\documentclass[conference]{IEEEtran}
\IEEEoverridecommandlockouts
\usepackage{cite}
\usepackage{amsmath,amssymb,amsfonts}
\usepackage{algorithm}
\usepackage{algorithmicx}
\usepackage{varwidth}
\usepackage{algpseudocode}
\usepackage{algorithm}
\usepackage{graphicx}
\usepackage{textcomp}
\usepackage{xcolor}
\usepackage{enumitem}
\usepackage{array}
\usepackage{multirow}
\usepackage{amsthm}
\usepackage{booktabs}
\def\BibTeX{{\rm B\kern-.05em{\sc i\kern-.025em b}\kern-.08em
    T\kern-.1667em\lower.7ex\hbox{E}\kern-.125emX}}

\begin{document}

\title{PICASSO: Unleashing the Potential of GPU-centric Training for Wide-and-deep Recommender Systems\\
\thanks{$^*$The first two authors contribute equally to this paper.}
}

\author{\IEEEauthorblockN{Yuanxing Zhang$^*$, Langshi Chen$^*$, Siran Yang, Man Yuan, Huimin Yi, Jie Zhang, Jiamang Wang, Jianbo Dong,\\
Yunlong Xu, Yue Song, Yong Li, Di Zhang, Wei Lin, Lin Qu, Bo Zheng}
\IEEEauthorblockA{Alibaba Group, China\\
\{yuanxing.zyx,langshi.cls\}@alibaba-inc.com}
}

\theoremstyle{plain}
\newcommand{\hbsys}{PICASSO}
\newcommand{\hsdl}{WDL}
\newcommand{\dfi}{WDL}
\newcommand{\hbworker}{PICASSO-Executor}
\newtheorem{problem}{Problem}

\maketitle

\begin{abstract}
The development of personalized recommendation has significantly improved the accuracy of information matching and the revenue of e-commerce platforms.
Recently, it has two trends: 1) recommender systems must be trained timely to cope with ever-growing new products and ever-changing user interests from online marketing and social network; 2) state-of-the-art recommendation models introduce deep neural network (DNN) modules to improve prediction accuracy. Traditional CPU-based recommender systems cannot meet these two trends, and GPU-centric training has become a trending approach.
However, we observe that GPU devices in training recommender systems 
are underutilized, and they cannot attain an expected throughput 
improvement as what it has achieved in Computer Vision (CV) and 
Neural Language Processing (NLP) areas.
This issue can be explained by two characteristics of 
these recommendation models: First, they contain up to a thousand of
input feature fields, introducing fragmentary and memory-intensive operations;
Second, the multiple constituent feature interaction submodules introduce substantial small-sized compute kernels. 
To remove this roadblock to the development of recommender systems, we propose a novel framework named \hbsys{} to accelerate the training of recommendation models 
on commodity hardware.
Specifically, we conduct a systematic analysis to reveal the bottlenecks encountered in training recommendation models.
We leverage the model structure and data distribution to unleash the potential of hardware through our packing, interleaving, and caching 
optimization.
Experiments show that \hbsys{} increases the hardware utilization by an order of magnitude on the basis of state-of-the-art baselines and brings up to 6$\times$ throughput improvement for a variety of industrial recommendation models.
Using the same hardware budget in production, \hbsys{} on average shortens the walltime of daily training tasks by 7 hours, significantly reducing the delay of continuous delivery.
\end{abstract}

\begin{IEEEkeywords}
Scalable training acceleration, Recommender system, Categorical data processing, Feature interaction
\end{IEEEkeywords}

\section{Introduction}
\label{sec:intro}

\begin{figure}[ht]
  \centering
  \includegraphics[width=\linewidth]{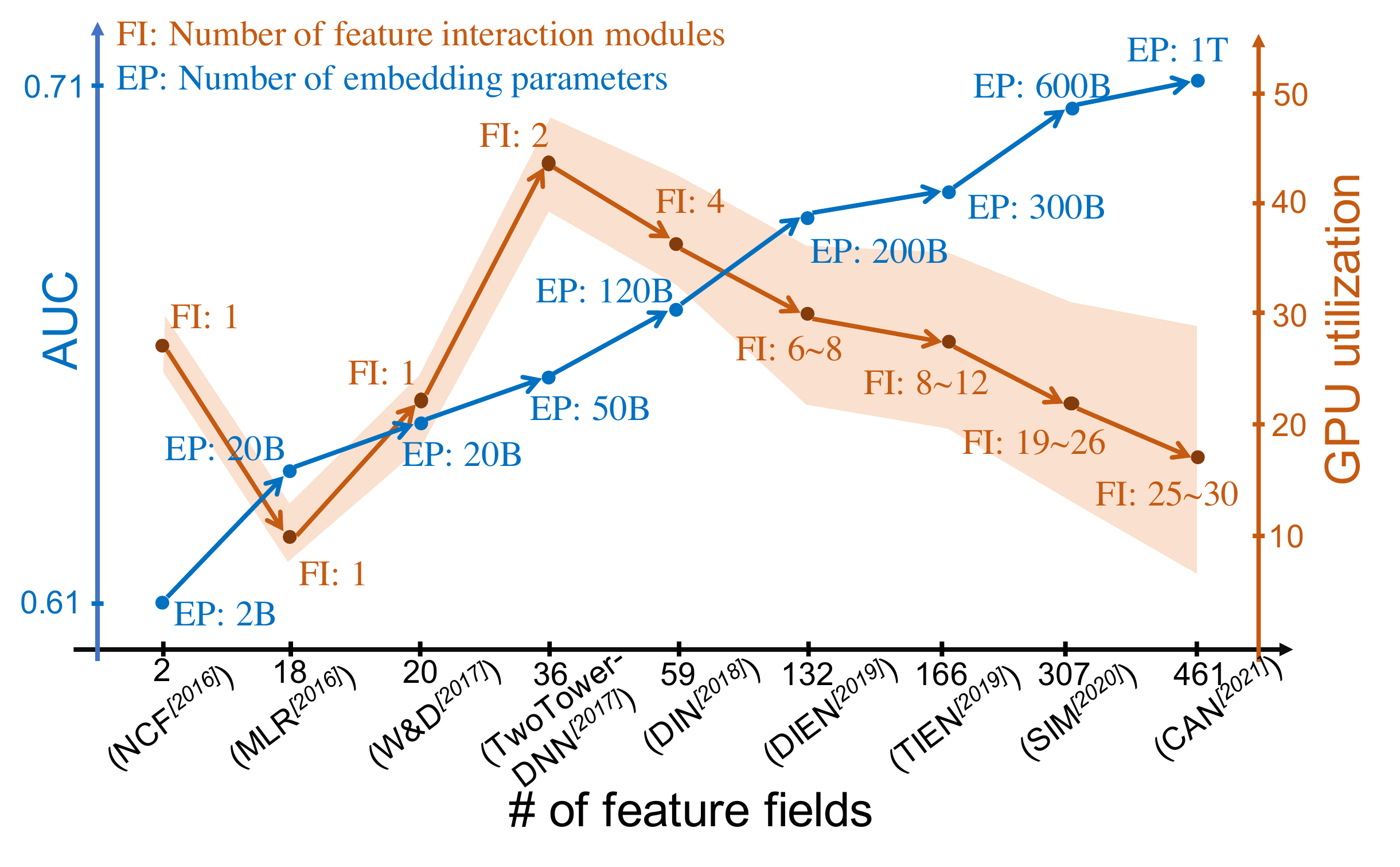}
  \caption{The trend of WDL approaches from the perspective of a business task in Alibaba. The average GPU utilization of these models is collected from the training workloads by an in-house optimized Tensorflow in PS training strategy. The statistics show that along with the performance improvement, training WDL models does not take full advantage of the GPU resources by the canonical training frameworks (which reach 95\%+ GPU utilization when training CV or NLP models of same scale).}\label{fig:teaser}
\end{figure}

Recommender systems have nowadays become the key for higher revenue, user engagement, and customer retention on social network and E-commerce platforms.
To cope with the explosive data growth, recommender systems are quickly evolving from 
collaborative filtering~\cite{schafer2007collaborative} (CF) to deep neural network (DNN) models and consistently improving the task accuracy, as shown in Fig.~\ref{fig:teaser}.
Starting from Google's Wide\&Deep~\cite{cheng2016wide}, the 
innovation of industrial-scale recommendation models~\cite{guo2017deepfm, zhou2018deep, zhou2019deep, naumov2019deep, qi2020search, zhou2020can}
follows two trends: 1) the embedding layer becomes wider, consuming up to thousands of 
feature fields~\cite{naumov2019deep}; 2) the feature interaction layer is going deeper~\cite{naumov2019deep, qi2020search, zhou2020can} by leveraging multiple
DNN submodules over different subsets of features. We denote these models as 
Wide-and-Deep Learning (\hsdl{}) Recommendation Models.

The industrial \hsdl{} models must be periodically re-trained to reflect user interest drift and new hot spots timely and accurately.
Hence, high training throughput is critical for \hsdl{} models to catch up streaming data and reduce the latency of continuous delivery~\cite{chen2015continuous}.
Training the state-of-the-art \hsdl{} models via 
Parameter Server (PS)~\cite{li2014scaling} on a large-scale distributed CPU cluster is 
time-consuming owing to insufficient computation capability to accomplish deep feature interactions.
Recent work, represented by Facebook's TorchRec\footnote{https://github.com/facebookresearch/torchrec, \emph{Accessed: 2021.11.30}}, Baidu's PaddleBox~\cite{MLSYS2020_f7e6c855},
and NVIDIA's HugeCTR\footnote{https://github.com/NVIDIA-Merlin/HugeCTR, \emph{Accessed: 2021.11.30}} prefer a GPU-centric 
synchronous training framework on \hsdl{} workloads because 
high-end NVIDIA GPU (e.g., NVIDIA Tesla V100) has a 30x higher single precision FLOPS over Intel CPU~\footnote{https://www.nvidia.com/en-us/data-center/v100/, \emph{Accessed: 2021.11.30}}.
Most of these efforts do improve the training 
throughput by leveraging GPU devices.
However, we observe substantial hardware underutilization (e.g., measured GPU utilization) along with the growing number of feature fields and feature interaction modules, as illustrated in Fig.~\ref{fig:teaser}.
This implies that the \hsdl{} training workloads are far from achieving the 
peak performance of hardware, and a further acceleration shall be 
anticipated. 
Although customizing hardware for a specific 
\hsdl{} workload pattern could be an option~\cite{smelyanskiy2019zion}, there are following concerns: 1) We have various \hsdl{} designs owning markedly different workload patterns (e.g., the number of feature fields, the submodules of feature interaction layer), and new \hsdl{} 
models are emerging monthly; 2) For public cloud usage, commodity hardware is preferred for the sake of budget and elasticity.
Thus, we raise two questions of \emph{What causes the hardware underutilization issue
in training \hsdl{} models?}, and \emph{Can we address this issue from software system's perspective?} 
We conduct a systematic analysis (detailed in \S\ref{sec:implication}) across a
variety of \hsdl{} workloads and obtain implications as follows:
\begin{enumerate}[leftmargin=*]
\item \hsdl{} model training has fragmentary operations within embedding and 
  feature interaction layers because of the massive number (up to thousands) of
  feature fields, which brings in a non-trivial overhead of launching operations 
  (e.g., issuing a CUDA kernel to CUDA streams) and hardware underutilization. 
\item The embedding layer of \hsdl{} model mainly consists of memory-intensive and 
  communication-intensive operations (in distributed environments), 
  while the feature interaction and multi-layer perceptron (MLP) own computation-intensive operations. The computing resource
  would be underutilized in processing the large volume of embedding parameters and leads to a pulse-like GPU usage.
\end{enumerate}
We then propose a novel framework with Packing, Interleaving and Caching Augmented Software System Optimization (\hbsys{}) to answer the above two questions.
First, we create fine-grained embedding feature groups. Operations within the same group are packed to reduce the number of fragmentary operations;
Second, operations from distinct groups are interleaved from both of data level and kernel level to improve hardware utilization;
Third, we develop a data distribution-aware caching mechanism leveraging the large volume of DRAM and the high bandwidth of GPU device memory. \par
Evaluation shows that \hbsys{} significantly improves the GPU utilization during training a variety of industrial \hsdl{} models, and it accelerates the throughput by an order of magnitude compared to state-of-the-art generic training frameworks.
\hbsys{} has been deployed in our in-house training cluster, appearing as \underline{XDL2} within Alibaba and \underline{HybridBackend} in AliCloud.
The delay to continuous delivery is decreased from 8.6 hours to 1.4 hours on average, which is unprecedented inside Alibaba and instructive to the community.
We summarize the main contributions of this paper as follows:
\begin{enumerate}[leftmargin=*]
  \item We analyze the hardware underutilization and reveal the cause when training \hsdl{} models by GPU-centric synchronous frameworks.
  \item We propose \hbsys{} that resolves the underutilization issue with a software-system approach that is applicable to commodity hardware. 
  \item We build a system that sustains our daily production workloads with up to a trillion parameters and petabytes of training data, achieving 6x training performance speedup
    on average without increasing the budget. \hbsys{} has been released for public cloud usage~\footnote{https://github.com/alibaba/HybridBackend}.
\end{enumerate}

\begin{figure}[tp]
  \centering
  \includegraphics[width=\linewidth]{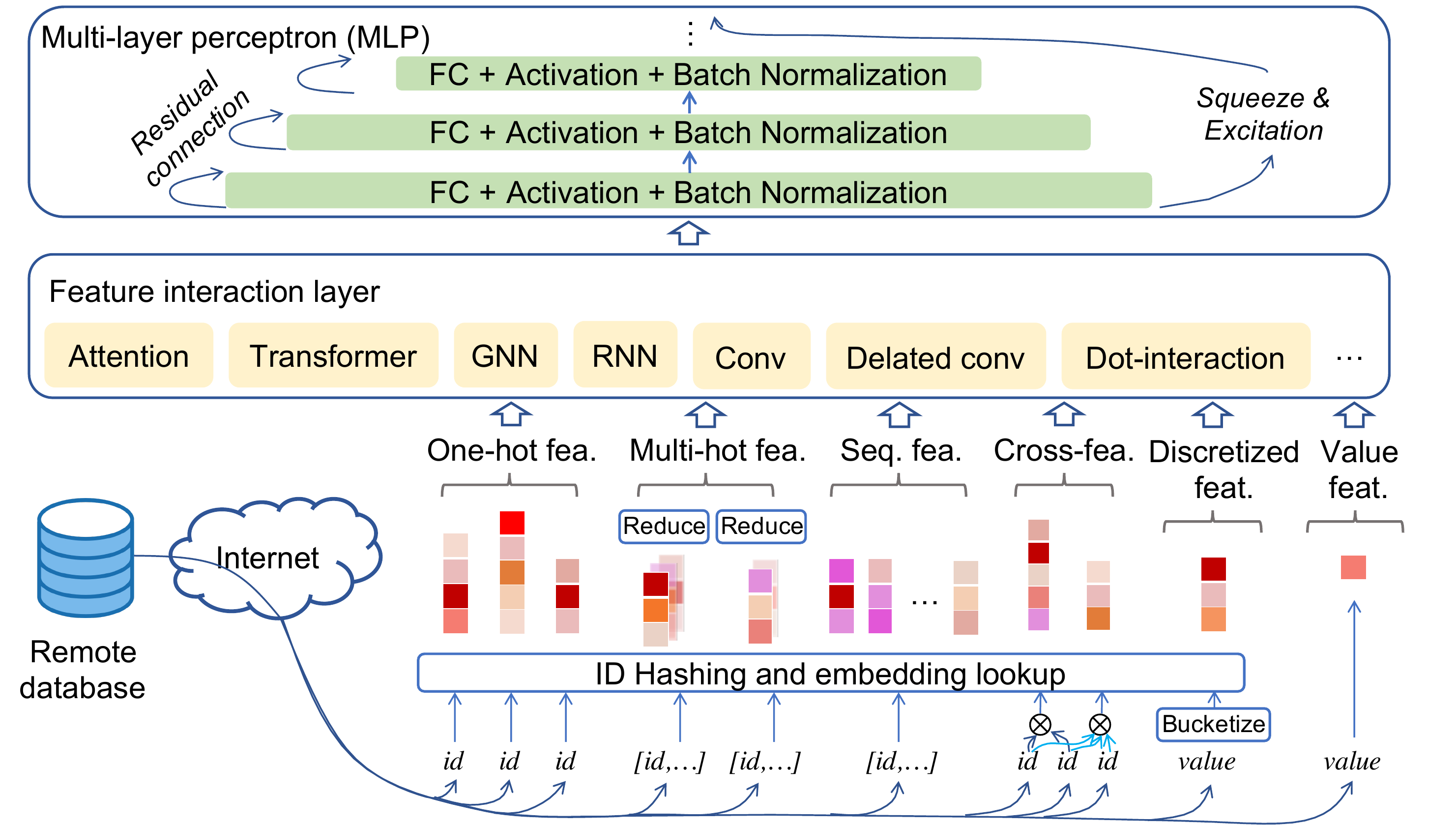}
  \caption{A Standard architecture of \hsdl{}, 1) \emph{The embedding layer} ingests multiple fields of features (as opposed to NLP tasks with only one field of ``word''), adopts various hashing or numerical operations, and queries embedding (in various dimension) from memory;
  2) \emph{The feature interaction layer} processes the fetched feature embeddings through a number of feature interaction modules; 3) the outputs of the modules are concatenated and fed to the multi-layer perception (MLP) for yielding final predictions.}\label{fig:rec_model}
\end{figure}

\section{Implication from Analyzing Workloads}
\label{sec:implication}


\subsection{Architecture of \hsdl{} Models}
\label{subs:implication-arch-models}

\hsdl{} models share a typical architecture as shown in Fig.~\ref{fig:rec_model}:
\noindent \textbf{Data Transmission Layer} processes streamed training data in the form of categorical feature identity (ID) as well as dense feature vectors. 
Categorical feature IDs usually have a varying length (i.e., multi-hot or non-tabular data), and they can reach up to tens or hundreds of MBytes within a batch. 
Normally, the data is stored in remote databases and requires a transmission via Ethernet;

\noindent \textbf{Embedding Layer} projects a high-dimensional feature space of sparse categorical features to a low-dimensional embedding feature space.
The embedding parameters are 
represented by dense vectors named 
\emph{feature embedding}, and they
are stored in DRAM in the form of \emph{embedding table}s. Each 
feature embedding can be queried 
by its categorical feature ID to 
get trained in \hsdl{}.
Since a large volume of feature embeddings would be queried from DRAM per training batch, the embedding layer 
is dominated by memory-intensive operations.
\begin{figure*}[t]
\centering
\begin{minipage}[t]{0.28\linewidth}
\centering
\includegraphics[width=\linewidth]{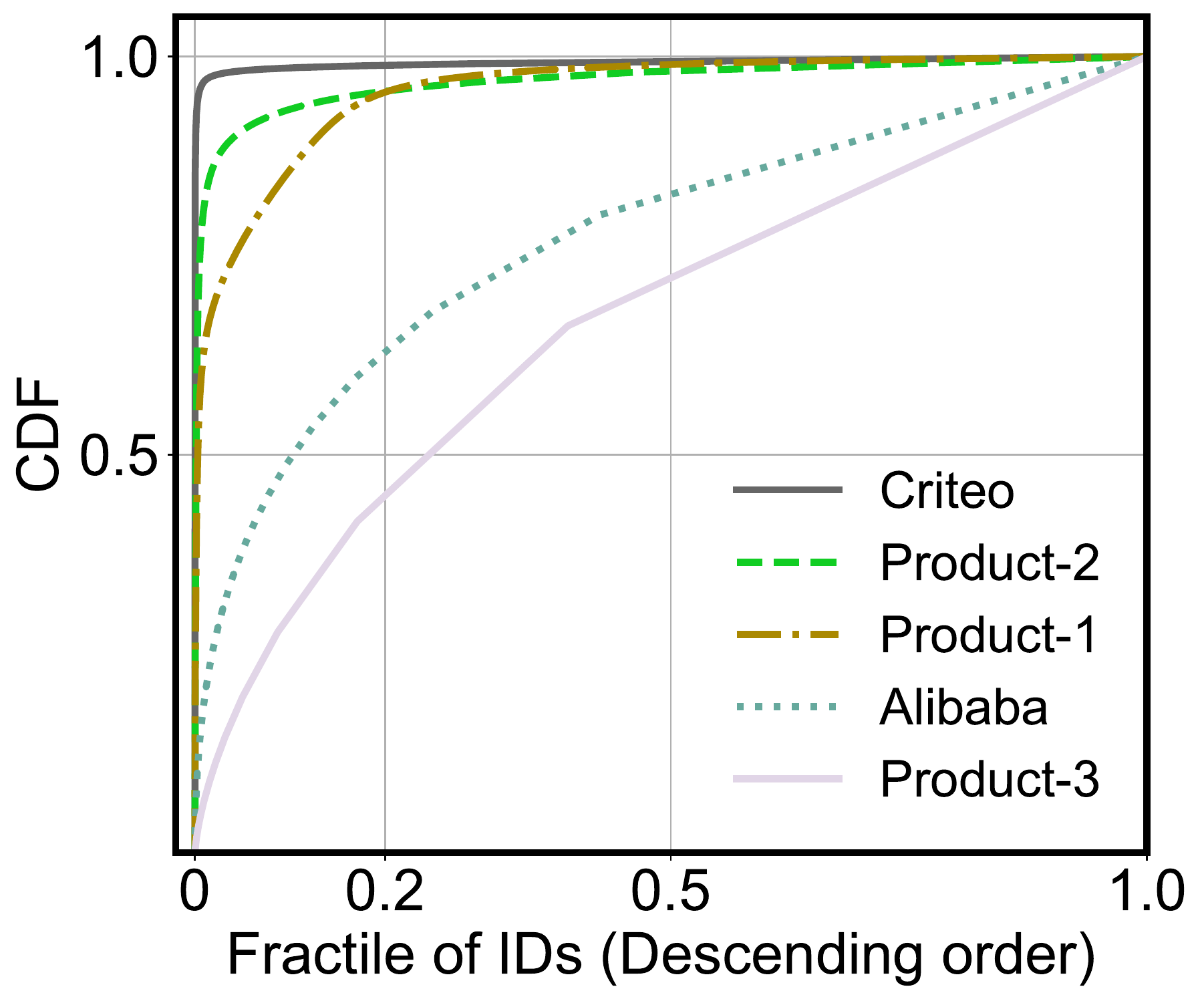}
\caption{Distribution of categorical feature IDs across representative \hsdl{} datasets.}
\label{fig:investigation}
\end{minipage}
\hfill
\begin{minipage}[t]{0.7\linewidth}
\centering
\includegraphics[width=0.99\linewidth]{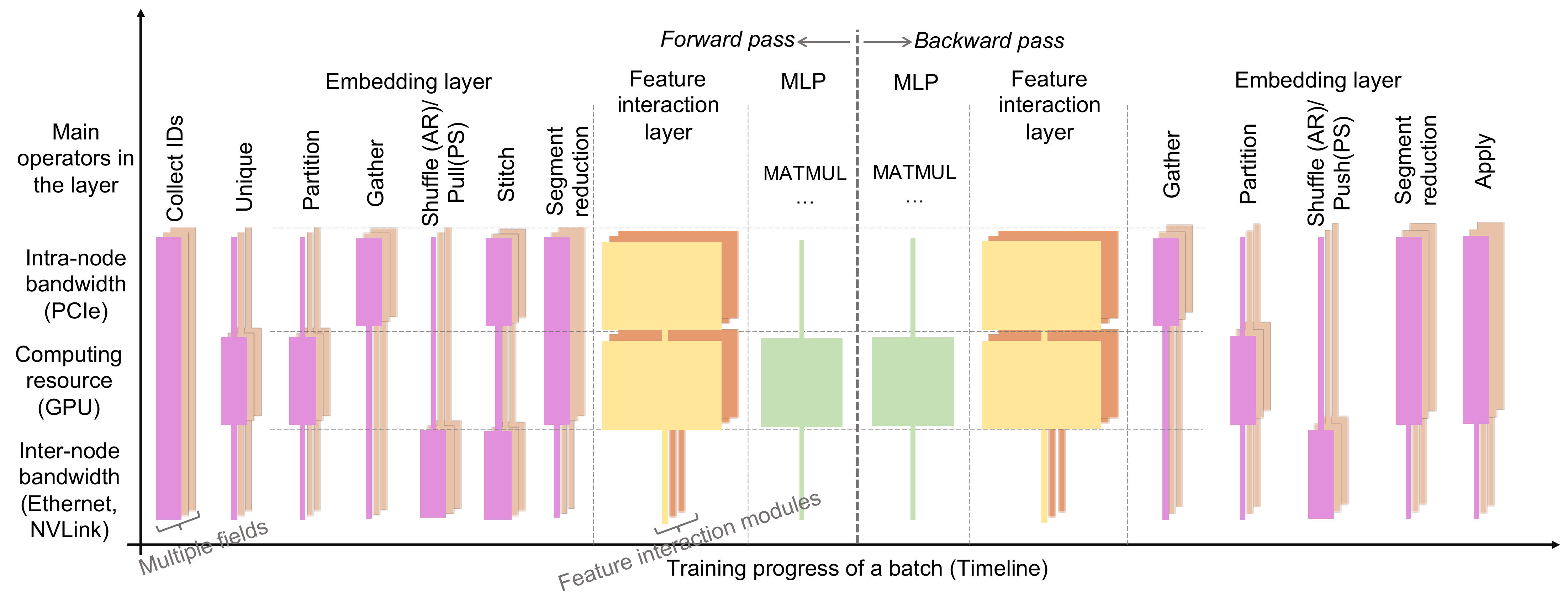}
\caption{Low-level projection of the standard architecture of \hsdl{} models: the multiple feature fields and constituent feature interaction modules would duplicate a large number of operations in the computation graph.}\label{fig:lowlevel}
\end{minipage}
\end{figure*}

\noindent \textbf{Feature Interaction Layer} would first organize feature embeddings from the embedding layer into several groups. Each group applies an individual \emph{feature interaction module}, such as 
GRU~\cite{chung2014empirical} and Transformer~\cite{vaswani2017attention}, to extract useful information from intra-group feature embeddings. 
The outputs of the constituent feature interaction modules are then concatenated to form a final output of the feature interaction layer.
There could be tens of constituent feature interaction modules to engage different subsets of feature embedding, yielding up to hundreds of thousands of operations.

\noindent \textbf{MLP} builds fully-connected layers to provide final predictions by the training data of a batch. 
MLP also contains computation-intensive architectural units such as batch normalization~\cite{ioffe2015batch} and residual connection~\cite{he2016deep}.
It is worth noting that the accuracy loss is usually intolerable in many business scenarios. Thus, generic acceleration strategies (e.g., half-precision, quantized pruning \cite{alistarh2017qsgd}, gradient staleness) may only apply to a fraction of \hsdl{} models.

\subsection{Data Distribution in \hsdl{} Workload}
\label{subs:data-distribution}

The categorical feature IDs of each feature field are usually skewed or non-uniformly distributed.
We investigate the data distribution of five
representative \hsdl{} datasets (statistics are listed in Tab.~\ref{tab:dataset}) and Fig.~\ref{fig:investigation}. 
When being sorted in a descending order by frequency, 20\% of IDs will cover 70\% by average and up
to 99\% of the training data. Therefore, it is beneficial to cache frequently accessed data when training \hsdl{} models. 

\subsection{Distributed Training Strategies of \hsdl{} Model}
\label{subs:implication-train-mode}

In general, three types of training strategies are 
adopted in training \hsdl{} models in a GPU-centric distributed system:

\noindent \textbf{Parameter Server (PS) Strategy} \cite{li2014scaling} is still the de facto training strategy applied in industry,
where training data is partitioned across multiple \emph{worker nodes} while model parameters are partitioned across multiple \emph{server nodes}.
Worker nodes would \emph{pull} down model parameters from server nodes and train them by using local partitioned training data;
By the end of each iteration, worker nodes would \emph{push} up
corresponding gradients back to server nodes asynchronously to 
update the parameters.

\noindent \textbf{Data-parallel (DP) Strategy} is the default distributed training strategy by frameworks such as Tensorflow~\cite{cheng2017tensorflow} and PyTorch~\cite{li2020pytorch}.
The training data is evenly partitioned across all worker nodes while the model parameters are replicated across all worker nodes. It uses a collective communication primitive named 
\emph{Allreduce} to aggregate gradients and thus updates local replica of model parameters synchronously.

\noindent \textbf{Model-parallel (MP) Strategy}~\cite{dean2012large,kim2019parallax} has no server nodes. Instead, it partitions and stores all parameters across multiple worker nodes.
It then uses a collective communication primitive 
named \emph{AllToAllv} to exchange data 
among all worker nodes synchronously.

\subsection{Characterization of \hsdl{} Workload}
\label{subs:characterize_model_arch}

We first conduct a low-level projection of \hsdl{} workloads to the underlying hardware, and we then summarize three representative workload patterns. \par

Each \hsdl{} layer consists of a set of \emph{operator}s 
from the algorithmic perspective.
An operator is normally implemented as a kernel in programs, and 
their invocations during the training are referred as the \emph{operations}.
The execution of an operation requires various hardware resources.
Fig.~\ref{fig:lowlevel} illustrates a low-level projection of a standard \hsdl{} model trained within a distributed system with respect to three types of hardware resources: 
intra-node bandwidth (e.g, DRAM and PCIe bandwidth), computing resource (e.g., GPU Streaming Multiprocessors (SM)), and inter-node bandwidth (NVLink and Ethernet bandwidth). 
In distributed training of \hsdl{} models, the embedding 
layer mainly consists of the following operators: \texttt{Unique} (eliminate redundant categorical feature IDs to 
reduce memory access overhead), \texttt{Partition} (partition categorical feature IDs into local IDs and remote IDs), \texttt{Gather} (query local IDs from embedding tables), \texttt{Shuffle} (communicate remote worker nodes to fetch feature 
embeddings belonging to remote IDs), \texttt{Stitch} (concatenate locally queried feature embeddings and remotely fetched
feature embeddings), and \texttt{SegmentReduction} 
(pooling feature embedding by segments, e.g., summation of 
behaviour feature embeddings from the same \emph{user}).
The embedding layer has a majority of operators bounded by a dominant type of hardware 
resource (e.g., the \texttt{Shuffle} operator is bounded by inter-node bandwidth).
The feature interaction layer and MLP have operators mainly bounded by computation resources. 
Similarly, the backward pass can be treated as a mirror image of the forward pass.
In a synchronous GPU-centric training system, this low-level projection indicates that the hardware resource usage is sporadic, 
which implies that, at a time, the training would be 
bounded by one type of hardware resource while the other types of resource are underutilized. \par
This projection reflects three characteristics of \hsdl{} workloads compared to CV and NLP workloads. 
\begin{itemize}[leftmargin=*]
  \item Embedding layer and feature interaction layer involve a massive number of small-sized operations (i.e., embedding of single feature field could involve hundreds of operations, and \hsdl{} workload would have to process 
    up to thousands of feature fields), which has a significant launch overhead. 
  \item The invocations of the same operator from the embedding layer by different feature fields would race for the same type of hardware resource at a time, which bounds the throughput when the hardware resource is relatively strained (e.g., PCIe bandwidth between CPU and GPU).
  \item The skewed data distribution as described in \S\ref{subs:data-distribution} would cause unbalanced hardware resource usage across worker nodes in distributed systems, which shall compromise the throughput during synchronous training.
\end{itemize}

\begin{figure}[tp]
  \centering
  \includegraphics[width=\linewidth]{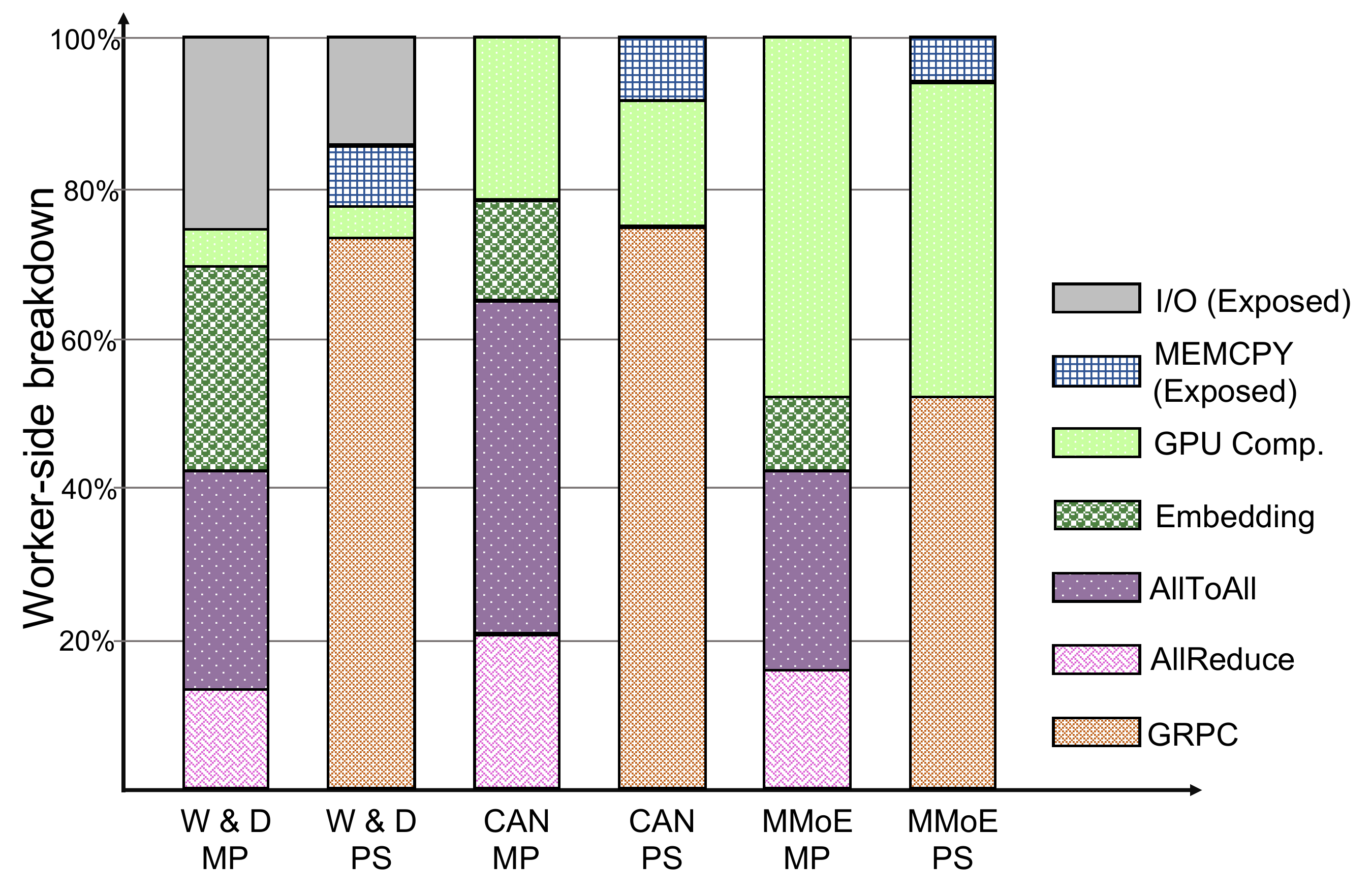}
  \caption{Worker-side breakdown of the three \hsdl{} models by PS and MP strategies (``exposed'' indicates the period when the operation blocks all the others).}\label{fig:breakdown}
\end{figure}

We probe into the statistics of \hsdl{} workloads from a cluster of NVIDIA Tesla V100s (``EFLOPS'' in Tab.~\ref{tab:testbed}) at Alibaba Cloud (composed of commodity hardware devices, as specified in \cite{dong2020eflops}). 
The models are implemented by Tensorflow under either PS strategy or MP strategy as introduced in \S\ref{subs:implication-train-mode}.
We observe three representative patterns from the profiling and worker-side performance breakdown, as shown in Fig.~\ref{fig:breakdown}.

\begin{itemize}[leftmargin=*]
  \item \emph{I/O\&Memory Intensive Workload}. \hsdl{} models represented by \emph{W\&D}~\cite{cheng2016wide} have
substantial data transmission and embedding lookup operations, where the I/O may not be fully overlapped by the other procedures, as shown in Fig.~\ref{fig:breakdown}.
I/O\&Memory intensive models emerge along with the prosperity of feature engineering and transferable pretrained embedding, where massive feature fields should be tackled to achieve the best accuracy.
Even with I/O optimization, the exposed I/O and memory access still 
occupy around 20\% of the total training walltime in Fig.~\ref{fig:breakdown}.
  \item \emph{Communication Intensive Workload}. This type of workload spends most of the time in communication-related operations.
The large volume of communication from fetching remote high-dimensional embedding features as well as the frequent parameter exchange from high-order cross features cause severe communication overhead in distributed \hsdl{} workloads.
We take CAN~\cite{zhou2020can} for instance, which is recently derived from 
DIN~\cite{zhou2018deep} and DLRM~\cite{mudigere2021high}. CAN contains a combination of feature interaction modules over a substantial number of feature fields, and therefore it brings up 
an extensive communication overhead by around 60\% in MP mode and 70\% in PS mode as shown in Fig.~\ref{fig:breakdown}.
  \item \emph{Computation Intensive Workload}. Some \hsdl{} models are throttled by computation operations, since deep and complex \hsdl{} models benefit from the advancement in the domain of CV and NLP.
A variant of MMoE~\cite{zhao2019recommending}, which is derived 
from canonical DIN~\cite{zhou2018deep} and owns 71 experts at MLP, 
serves for scenario-aware CTR prediction in our business. In Fig.~\ref{fig:breakdown}, MMoE spends about 50\% of the training time in arithmetic calculations.
In practice, computation-intensive \hsdl{} models tend to work for scenarios of multiple sub-tasks \cite{sheng2021one} (e.g., multi-objective learning, meta-learning), super-complicated computation \cite{senior2020improved} (e.g., extremely-deep network over a number of feature fields), and multi-modal co-training \cite{wei2019mmgcn}.
\end{itemize}

\section{System Design of \hbsys{}}%
\label{sec:design_and_architecture}

In this section, we first introduce the overall architecture of \hbsys{}, which accommodates the hardware topology of GPU-centric training cluster and thus allows a hybrid distributed training strategy 
for \hsdl{} models. Second, we introduce a three-fold key idea of accelerating training by improving hardware utilization.

\subsection{Hybrid Distributed Training Strategy}%

\begin{figure}[t]
    \centering
    \includegraphics[width=1.0\linewidth]{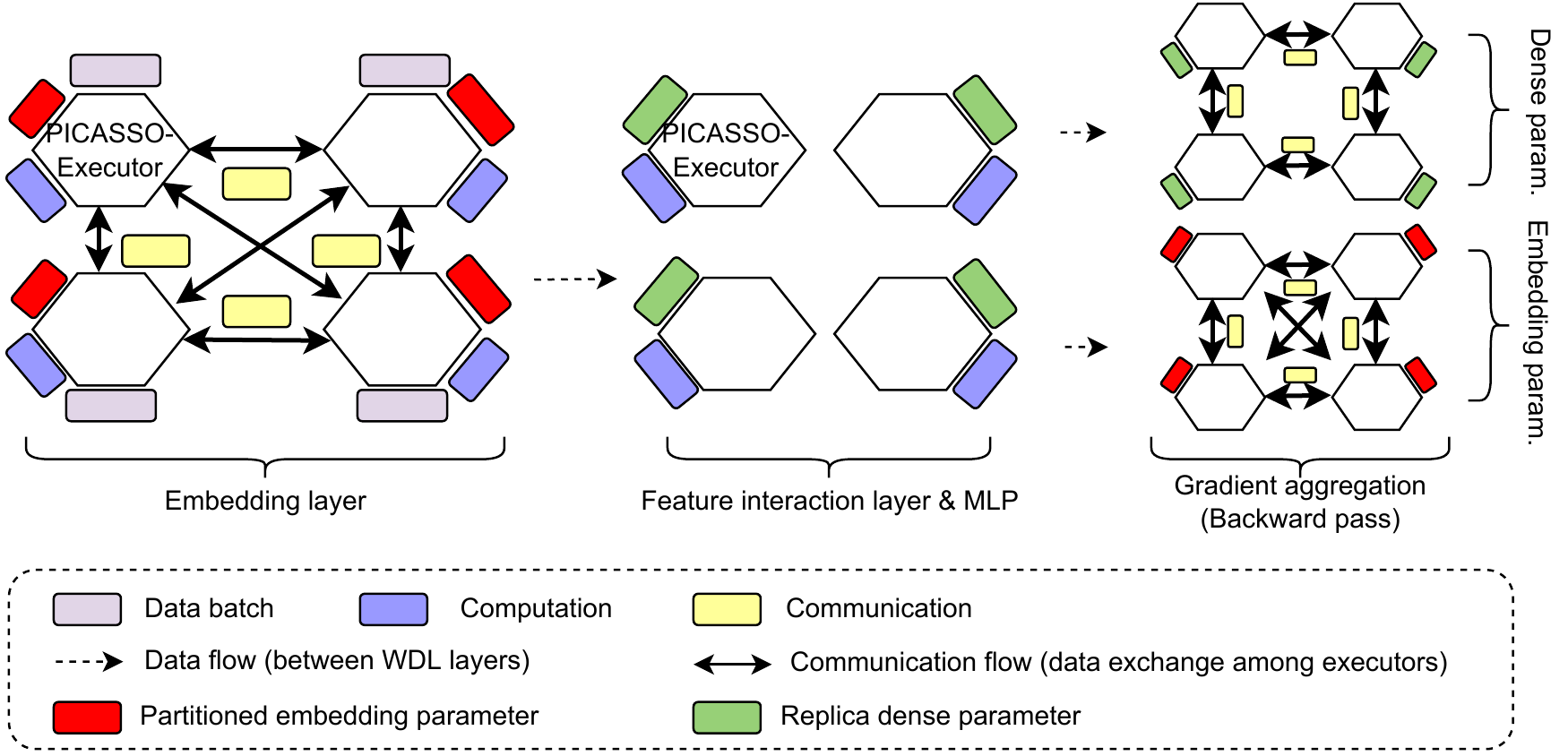}
    \caption{\hbsys{} allows the hybrid strategy of MP and DP, where embedding parameters are partitioned across \hbworker{}s (MP), and dense parameters are replicated across \hbworker{}s (DP). Besides, canonical DP and MP are also supported by \hbsys{}.}%
    \label{fig:sys-design-hb}
\end{figure}

We propose the architecture of \hbsys{} in Fig.~\ref{fig:sys-design-hb}, which is designed for GPU-centric clusters. One such cluster usually consists of multiple homogeneously configured
\emph{machines (cluster node)}, and each machine instead has a heterogeneous architecture, including processors (e.g., Intel CPU), accelerators (e.g., NVIDIA GPU), 
and the uncore system (e.g., DRAM). In addition, there are interconnects like PCIe and NVLink among components within a machine, and all these machines are 
further connected through Ethernet as a distributed system. Correspondingly, \hbsys{} sets up multiple \emph{\hbworker{}}s, which are mapped to different machines in the cluster. 
Each \hbworker{} has heterogeneous hardware resource: 1) GPU Stream Multiprocessor (SM) and CPU physical cores as computation resource; 2) hierarchical memory subsystem 
made of GPU device memory, DRAM, Intel Persistent Memory, and SSD (if accessible) as memory storage resource; 3) hierarchical interconnects like NVLink, PCIe, InfiniBand, and Ethernet as communication resource. \par

With this architecture, \hbsys{} can customize hybrid distributed training strategies for different layers within \hsdl{} model as follows:
\begin{itemize}[leftmargin=*]
  \item The embedding layer has a tremendous volume of embedding parameters, and it adopts a model-parallel (MP) strategy. 
    The embedding parameters are partitioned across all \hbworker{}s and stored within the hierarchical memory subsystem.
    The parameters are exchanged synchronously among \hbworker{}s via 
    the AllToAllv collective communication primitive.
  \item Feature interaction layer and MLP have a much smaller volume of parameters than the embedding layer. 
    We adopt a data-parallel (DP) strategy for those two layers, where the parameters are replicated across all \hbworker{}s and aggregated via the Allreduce primitive.
\end{itemize}

\subsection{Packing}
\label{subs:packing}

\begin{figure}[htpb]
    \centering
    \includegraphics[width=1.0\linewidth]{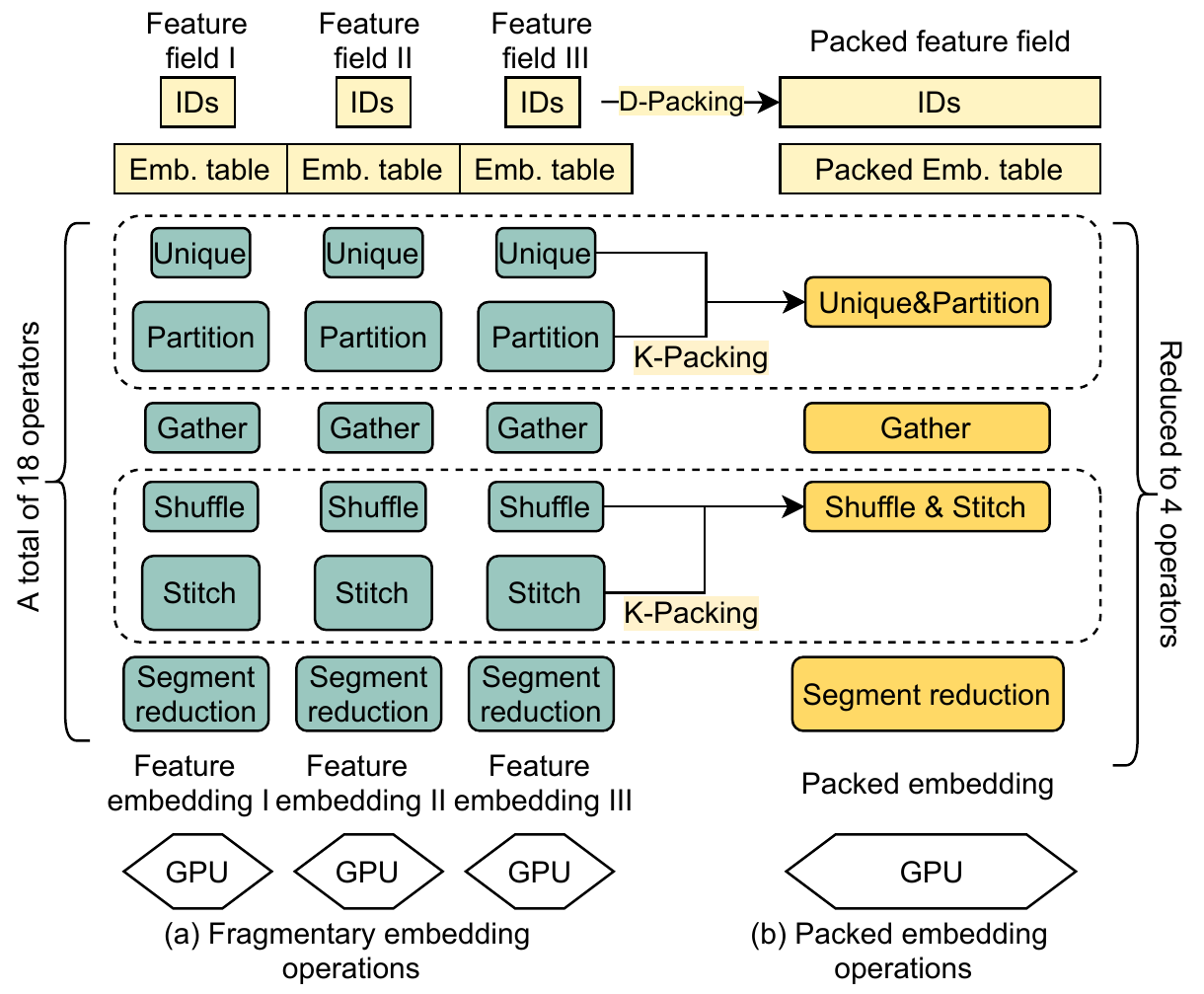}
    \caption{Illustration of \emph{packing} optimization:
    a) Without packing, looking up feature embedding from three embedding tables requires a total of 18 operators;
    b) By D-packing and K-packing, \hbsys{} reduces the total number of operators to 4 packed operators.}
    \label{fig:packing}
\end{figure}

To address the fragmentary operations as described in \S\ref{subs:characterize_model_arch}, we propose a \emph{Packing} method that effectively reduces the 
number of operations from two aspects:

\begin{figure*}[htp]
    \centering
    \includegraphics[width=\linewidth]{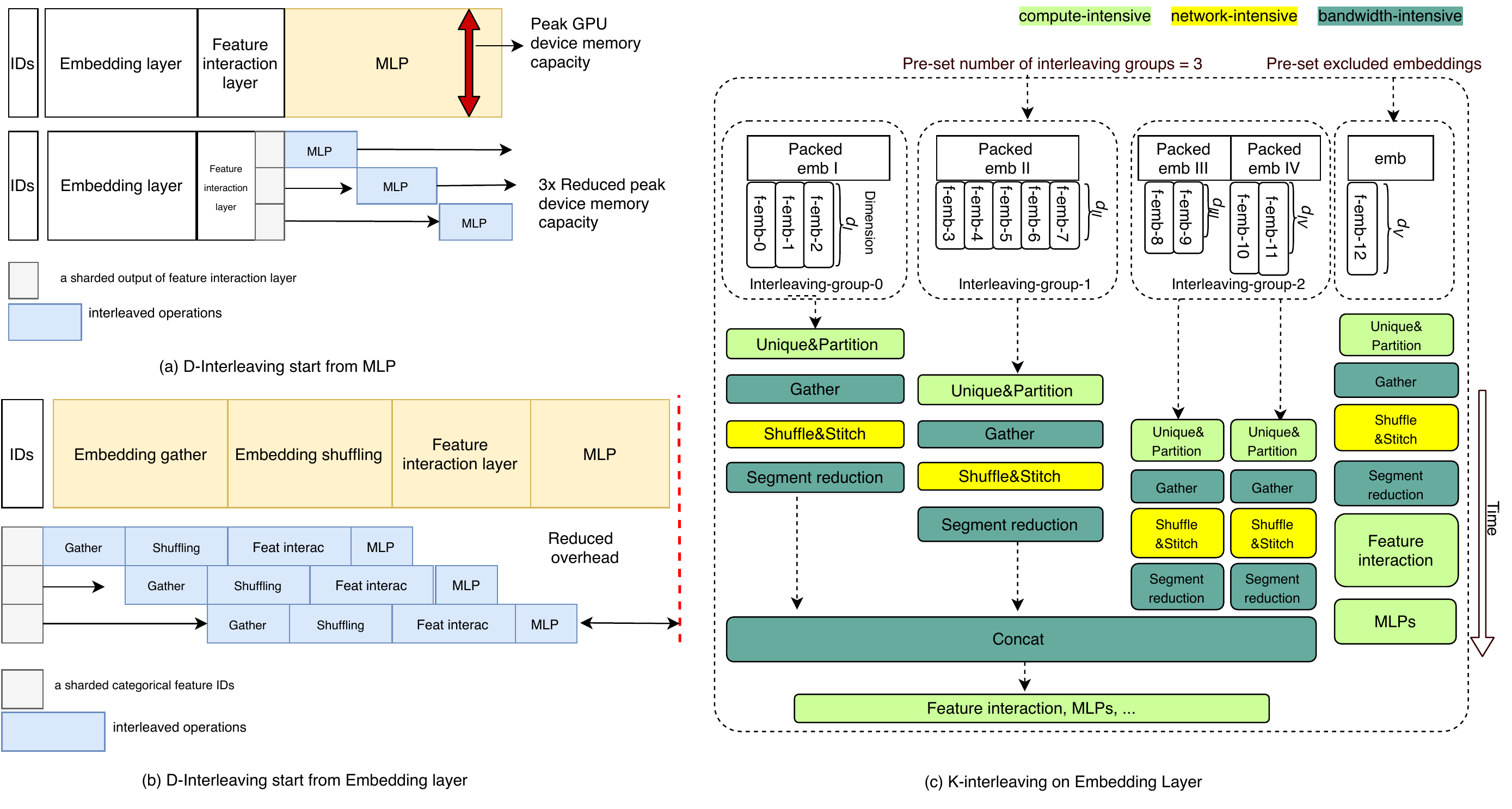}
    \caption{
     (a) \emph{D-Interleaving} starts from MLP to eliminate GPU device memory capacity
     bounds;
     (b) \emph{D-Interleaving} starts from the embedding layer to reduce 
     overhead.
     (c) \emph{K-Interleaving} the packed and non-packed embedding operations on different hardware resource.}%
    \label{fig:interleave}
\end{figure*}

\noindent\textbf{Data-Packing} (D-Packing). 
When categorical feature IDs from different feature fields are fed into the same operator within the embedding layer, 
\hbsys{} combines categorical feature IDs into a single \emph{packed ID tensor}. Thus, we can launch a single operation 
(named \emph{packed operation})
to process the packed data, which fits 
into the Single-Instruction-Multiple-Data (SIMD) 
programming paradigm~\cite{patterson2016computer} of NVIDIA GPU devices. Also, it significantly reduces the overhead 
of launching a massive number of operations onto GPU devices.

In addition, a naive strategy of packing up all 
categorical feature IDs into a single tensor may cause severe throughput issues.
For instance, industrial-scale recommender systems usually
implement embedding tables by using a \emph{hashmap} to accommodate 
the growing amount of feature embedding. 
A massive number (million-level) of concurrent
querying requests 
shall suffer from the low-level locks
of a hashmap. Therefore, we pack up categorical feature IDs 
when their embedding tables share the same feature dimension.
Thus, we obtain packed operations with the number proportional to 
that of distinct feature dimensions. Nevertheless, some packed
operations may still encounter too many concurrent queries to 
compromise the throughput of the hashmap due to 
skewed data distribution and large feature dimensions.  
We propose a 
method to evaluate the execution cost by estimating 
the parameter volume 
(the number of floats) within packed operations 
as shown in Equation~\ref{eq:pack-vars}.
\begin{equation}
  \label{eq:pack-vars}
  \text{CalcVParam}(T) = N\sum_{t\in T}(t_{\text{dim}}\sum_{\text{ID}\in t}\text{ID}_{\text{freq}}),
\end{equation}
where $N$ refers to the total number of categorical feature IDs, 
$T$ refers to a packed embedding table, $t_{\text{dim}}$ denotes the feature
dimension of an embedding table, and $\text{ID}_{\text{freq}}$ refers to 
the frequency of a categorical feature ID. 
$N$ and $\text{ID}_{\text{freq}}$s would be obtained from the statistics in the warm-up iterations.
If a packed operation has a high $\text{CalcVParam}(T)$ above average, we shall further evenly split it 
into multiple shards. For instance, suppose that we have one packed operation for 
all embedding tables with a feature dimension of 8 and the data distribution is uniform.
For embedding tables with a dimension of 32, 
we will create four shards of packed 
operations, each with a quarter of these embedding tables.

\noindent\textbf{Kernel-Packing} (K-Packing). Kernel fusion is already a widely adopted optimization for deep learning systems. There are mainly two approaches: 1) hand-written huge kernels; 2) 
compilation-based code generation. Huge kernels, such as fusing the whole embedding layer into a single CUDA kernel, would miss the opportunity of interleaving operations bounded by 
distinct hardware resources (see details in \S\ref{subs:interleaving}).
Compilation-based code generation relies on static input and output shapes from each operator to 
infer the suitable sizes for generated kernels. However, categorical feature IDs induce dynamic 
operator shapes that disrupt the efficiency of compilation techniques like Tensorflow XLA.
In contrast, our kernel-packing evaluates all kernels by their hardware resource 
utilization, and they are grouped into computation-intensive kernels, memory-intensive kernels, and 
communication-intensive kernels. We only fuse up kernels from the same kernel group and leave 
opportunities for interleaving their execution across different kernel groups.\par

Fig.~\ref{fig:packing} illustrates the process of our packing optimization. 
Categorical feature IDs are first grouped together (i.e., D-Packing).
Each group of categorical feature IDs is fed into a fused kernel (i.e., K-Packing) 
named \texttt{Unique\&Partition} to eliminate memory access and data communication 
from redundant IDs. The categorical feature IDs would be fetched from their 
local partition of embedding tables. We then develop another fused kernel named
\texttt{Shuffle\&Stitch} to fulfill a stitched output of the shuffle kernel and 
remove the explicit stitch kernel.

\subsection{Interleaving}
\label{subs:interleaving}

After applying the Packing optimization, we further develop two types of 
\emph{Interleaving} optimization to improve utilization across different 
hardware resources. 

\noindent\textbf{Data-Interleaving}(D-Interleaving): When \hsdl{} model is trained with
large batch sizes (e.g., tens of thousand), operations
would suffer from various hardware bounds. For instance, the footprint of GPU 
device memory from intermediate tensors (the so-called \emph{feature map} in Tensorflow) 
is proportional to the data batch size. Because the capacity of GPU device memory is 
restricted (e.g., 32GBytes in NVIDIA Tesla V100), large batch size is likely to cause 
an out-of-memory (OOM) issue and crash the training. 
However, large batch size is usually desired for both high accuracy and throughput in \hsdl{} training.
Therefore, \hbsys{} adopts a micro-batch-based 
data interleaving (D-interleaving) approach that allows users to slice
and interleave workloads starting from a specified layer of \hsdl{} models. 
To address the OOM issue of GPU device memory, we can 
divide the output embedding of feature interaction layer into several micro batches 
and apply D-Interleaving on MLP, 
where the peak GPU device memory usage can be amortized as shown in Fig.~\ref{fig:interleave} (a).
Also, we can divide categorical feature IDs into several micro batches and 
apply D-Interleaving to the rest of the training Fig.~\ref{fig:interleave} (b).
By default, we evenly divide data into micro batches to attain a load balancing, and the 
micro-batch size can be estimated by:
\begin{equation}
  \label{eq:micro-batch}
  \text{BS}_{\text{micro}} = \min_{\text{op} \in \text{layer}}(\text{RBound}_{\text{op}}/\text{RInstance}_{\text{op}}),
\end{equation}
where $\text{BS}_{\text{micro}}$ is the estimated micro-batch size, $\text{RBound}_{\text{op}}$ denotes 
the bound value of an operator's 
dominant hardware resource (e.g., GPU device memory capacity), and $\text{RInstance}_{\text{op}}$ denotes 
the cost per data instance from an operator's dominant hardware resource. Since the shape of 
an operator from the embedding layer is usually dynamic, no analytical values from 
Equation~\ref{eq:micro-batch} can be deduced in advance. Instead, 
we determine their values empirically or experimentally from warm-up iterations of training. 

\noindent\textbf{Kernel-Interleaving}(K-Interleaving): The Packing mechanism
transforms hundreds of operations into a small number of packed operations.
However, the packed operations from different embedding tables 
still race for the same hardware resource. For instance, when all the \texttt{Shuffle\&Stitch} 
operations are launched concurrently, the Ethernet bandwidth would 
throttle the training throughput. We propose a \emph{Kernel-Interleaving} (K-Interleaving) optimization that
establishes control dependencies among groups of packed operations as shown in Fig.~\ref{fig:interleave}. To ensure that a so-called 
\emph{Interleaving group} would not be bounded by various hardware
resources, we first determine the capacity of each interleaving group, denoted as $\text{Capacity}_{\text{g}}$, in terms of processed parameters by:
\begin{equation}
  \label{eq:capacity-i-group}
  \text{Capacity}_{\text{g}} = \min_{\text{op} \in \text{layer}}(\text{RBound}_{\text{op}}/\text{RParam}_{\text{op}}),
\end{equation}
where $\text{RBound}_{\text{op}}$ has 
the same definition as in Equation~\ref{eq:micro-batch} and
$\text{RParam}_{\text{op}}$ denotes the cost of training a 
parameter from an operator's dominant hardware resource.
Here, we simply treat the parameter volume as the cost in embedding lookup and exchange.
We could also change the number of interleaving groups by proportionally modifying $\text{Capacity}_{\text{g}}$ in \hbsys{}.
It is worth noting that we allow users to specify a 
\emph{preset excluded embedding}, where the packed operations have no control dependencies on the other K-Interleaving groups.
For instance, when the output (feature embedding) of some operations will not be 
concatenated with other feature embedding for the downstream layers, K-Interleaving can advance their downstream operations.

\subsection{Caching}
\label{subs:caching}

\begin{figure}[t]
    \centering
    \includegraphics[width=0.9\linewidth]{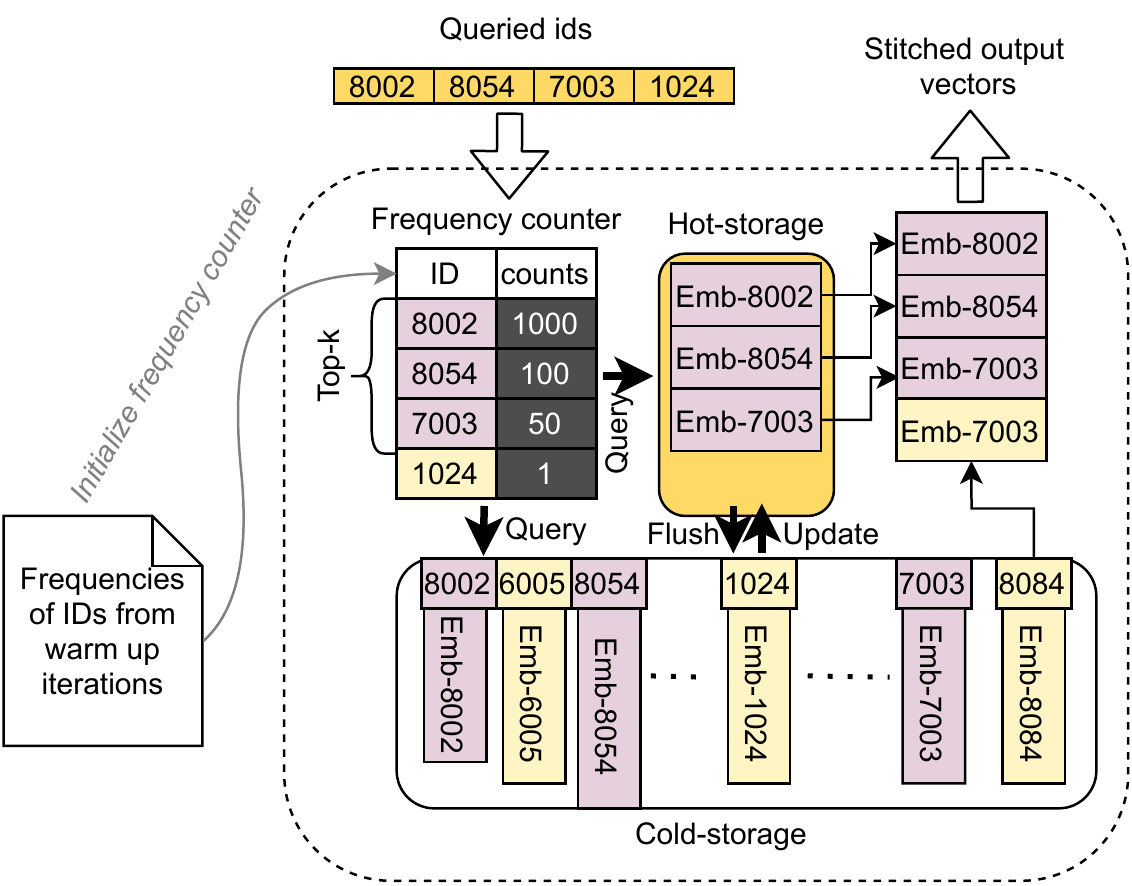}
    \caption{Gather embedding vectors (e.g., ``Emb-8002'') via \emph{Cache} approach upon \emph{HybridHash}.}%
    \label{fig:cache-and-weave-opt}
\end{figure}

\emph{Caching} is a widely applied system technique that utilizes the hierarchical memory subsystem to reduce memory access latency. 
However, the effectiveness of caching, i.e., the \emph{cache hit ratio}, depends on multiple factors such as the data distribution and the access pattern. 
In \S\ref{subs:data-distribution}, we observe that only 20\% of the categorical feature IDs 
are being queried in a high frequency, which motivates us to propose an optimization named 
\emph{HybridHash}. It aims to eliminate two hardware bounds: 1) The DRAM has a 
large capacity but is bounded by the memory access bandwidth; 2) The GPU device memory has a 
high bandwidth but is bounded by the limited capacity. \par

As shown in Fig.~\ref{fig:cache-and-weave-opt}, HybridHash
serves as a hashmap to store, fetch, and update embedding parameters. 
We refer to GPU device memory as the \emph{Hot-storage} and DRAM as the \emph{Cold-storage}. Unlike other GPU-based hashmap solutions, we consider
Hot-storage as an expensive resource, which shall avoid any waste of its capacity.
Therefore, we place the hashmap, a sparse data structure, on Cold-storage
and utilize Hot-storage only as a scratchpad to store and update the 
frequently accessed embedding. 
Recall from \S\ref{subs:data-distribution} that categorical feature IDs in \hsdl{} workload follow a 
certain distribution. It is reasonable to record the frequency of 
each queried ID from the hashmap during a predefined number 
of warm-up iterations. HybridHash then periodically 
loads the top-$k$ ($k$ would be determined by the size of Hot-storage) frequent embedding from Cold-storage to 
Hot-storage to retain the hottest categorical feature IDs.
After the warm-up steps, the majority of ID queries are
expected to hit in Hot-storage, and missed queries can be
handled by Cold-storage.
Note that HybridHash would place all data on 
Hot-storage when its capacity is found to be far beyond the total size of embedding tables during the warm-up steps. In addition, HybridHash can be extended to a multiple-level cache system, including devices like Intel's persistent memory and SSD.  
The algorithm of HybridHash is shown in Algorithm~\ref{alg:hybridhash}. 
L\ref{line:warm_start}-\ref{line:warm_end} depict the steps to collect statistics during the warmup iterations, L\ref{line:normal_start}-\ref{line:normal_end} introduce the rules to get embedding, and L\ref{line:update_start}-\ref{line:update_end} define the procedures to update the content in Hot-storage.
\begin{algorithm}
    \caption{Algorithm of HybridHash}
    \label{alg:hybridhash}
    \begin{algorithmic}[1]
        \Procedure{HybridHash}{IDs, itr}
        CStore: cold storage to hold hashmap \\
        HStore: hot storage as a cache \\
        FCounter: a host-side counter to record ID's frequency \\
        warmup\_iters: iterations to warm up Hybridhash  \\
        flush\_iters: flush HStore by top features in CStore every flush\_iters \\
        IDs: categorical feature IDs to query \\
        itr: current iteration 
        \If{$\text{itr} < \text{warmup\_iters}$}
        \For{$\forall id \in \text{IDs}$}\label{line:warm_start}
        \State $\text{FCounter}(id) \gets \text{FCounter}(id) + 1$
        \State $feat(id) \gets \text{CStore}(id)$
        \EndFor \label{line:warm_end}
        \Else
         \For{$\forall id \in \text{IDs}$} \label{line:normal_start}
         \If{$id$ is found in HStore}
         \State $feat\_hot(id) \gets \text{HStore}(id)$
         \Else
         \State $feat\_cold(id) \gets \text{CStore}(id)$
         \EndIf
         \State $\text{FCounter}(id) \gets \text{FCounter}(id) + 1$
         \EndFor 
         \State $feat = feat\_hot \cup feat\_cold$ \label{line:normal_end}
         \If{$\text{itr} \% \text{flush\_iters} = 0$} \label{line:update_start}
         \State $hot\_ids \gets \text{top-}k(\text{FCounter})$
         \State $\text{HStore} \gets \text{CStore}(hot\_ids)$ 
         \EndIf \label{line:update_end}
        \EndIf \\
        \Return $feat$
        \EndProcedure
    \end{algorithmic}
\end{algorithm}

\section{Experiment}
In this section, we conduct extensive experiments to answer the following research questions:
\begin{itemize}[leftmargin=*]
    \item \textbf{RQ1}: To what extent does \hbsys{} improve training throughput by unleashing the hardware potential compared to the state-of-the-art generic frameworks with computation optimizations?
    \item \textbf{RQ2}: How do the software system optimizations in \hbsys{} affect the utilization of each hardware resource?
    \item \textbf{RQ3}: How does \hbsys{} perform on the diversified \hsdl{} model architectures and feature fields?
\end{itemize}

\subsection{Experimental Setup}
\label{sub:exp_setup}

We conduct the experiments on \hbsys{} from two aspects: 
1) Benchmarking the performance of \hbsys{} with state-of-the-art frameworks 
on public datasets. 2) Evaluating the design of \hbsys{} by production-ready datasets and three representative models.
Tab.~\ref{tab:testbed} summarizes our testbeds, including public-accessible machines from AliCloud (Gn6e) for the performance benchmarking, as well as 
an on-premise cluster of Tesla-V100 (EFLOPS) for the system design evaluation. 
\begin{table}[t]
    \centering
    \footnotesize
    \caption{Specification of testbeds (per node)}
    \label{tab:testbed}
    \resizebox{\linewidth}{!}{
    \begin{tabular}{p{0.9cm}|p{1.3cm}p{1.7cm}p{1.1cm}p{1.4cm}}
    \toprule
    \textbf{Cluster} & \textbf{CPU} & \textbf{GPU} & \textbf{DRAM} & \textbf{Network} \\
    \midrule
    Gn6e & Xeon 8163 (96 cores) & 8xTesla V100-SXM2 (256GB HBM2) & 724GB DDR4 &32Gbps (TCP)   \\
    EFLOPS & Xeon 8269CY (104 cores) & 1xTesla V100S-PCIe (32GB HBM2) & 512GB DDR4 & 100Gbps (RDMA)\\
    \bottomrule
    \end{tabular}
    }
\end{table}

\noindent\textbf{Testing Models and Datasets}.
\emph{DLRM} \cite{mudigere2021high} is a benchmarking model proposed by Facebook and 
adopted by MLPerf; \emph{DeepFM} \cite{guo2017deepfm}, derived from Wide\&Deep model, is widely 
applied in industrial recommender systems; \emph{DIN} \cite{zhou2018deep} and \emph{DIEN} 
\cite{zhou2019deep} are two models training multi-field categorical data with complicated
feature interaction modules. We also utilize the three representative models
discussed in \S\ref{sec:implication} for a system-design evaluation. \par 

For benchmarking datasets, we collect: 1) \emph{Criteo} \cite{criteo}, 
a widely adopted click-through-rate (CTR) dataset by Kaggle and MLPerf 
\cite{mlperf}, and 2) \emph{Alibaba}\cite{qi2020search}, an open-sourced
industrial-level CTR dataset. 
For a system-design evaluation, we use in-house production datasets at
Alibaba, which has a large number of one- or multi-hot categorical features.
The statistics of these datasets are depicted
in Tab.~\ref{tab:dataset}.
The datasets are placed on a remote server to download via network.
Following common industrial settings, the models would go through 
only one epoch of the entire dataset and
adopt a full-precision training to avoid accuracy loss.

\begin{table}[t]
    \centering
    \caption{Statistics of datasets in experiments (lengths of sequential features are shown in parentheses)}
    \label{tab:dataset}
    \resizebox{\linewidth}{!}{
    \begin{tabular}{p{1.3cm}|p{0.95cm}p{1.22cm}p{2cm}|p{1.2cm}p{0.9cm}p{0.7cm}}
    \toprule
     \textbf{Dataset} & \textbf{Instances} & \textbf{\# numeric features} & \textbf{\# sparse feature fields} & \textbf{Model} & \textbf{Emb. dim.} & \textbf{\# of params} \\
    \midrule
    Criteo & 4B & 13 & 26 & DLRM/ DeepFM & 128 & 6B \\
    Alibaba & 13M & 0 & 1,207 (7+12$\times$100) & DIN/ DIEN & 4 & 6B \\
    \midrule
    Product-1 & Infinite & 10 & 204 & W\&D & 8$\sim$32  & 160B \\
    Product-2 & Infinite & 0 & 1,834 (334+30$\times$50) & CAN & 8$\sim$200  & 1T \\
    Product-3 & Infinite & 0 & 584 (84+10$\times$50) & MMoE & 12$\sim$128  & 1T \\
    \bottomrule
    \end{tabular}
    }
\end{table}

\noindent\textbf{State-of-the-art Training Frameworks}.
We evaluate and compare the performance of \hbsys{} with mainstream 
open-sourced \hsdl{} training frameworks, including: 
\emph{Tensorflow-PS} (abbr. TF-PS) of version 1.15 \cite{tf15} with an asynchronous PS
training strategy (one PS on CPU and multiple workers on GPUs). NVLink does not work in this training mode;
\emph{PyTorch} of version 1.8 \cite{pytorch} with a hybrid training 
strategy on \hsdl{} models with AllToAll communication (over NCCL)
developed by Facebook. The embedding tables towards different feature fields are manually placed on different GPUs based on their sizes;
\emph{Horovod} \cite{sergeev2018horovod} on PyTorch distributed
data-parallel (DDP) mode with an Allreduce communication.

\noindent\textbf{Evaluation Metrics}.
A group of metrics are employed in our experiments for comprehensive measurements:
\begin{itemize}[leftmargin=*]
    \item \emph{AUC} is a standard CTR metric to 
    evaluate the accuracy;
    \item \emph{Performance} refers to the 
    throughput of training system (instances per second per node (IPS)) and
    training walltime (GPU core hours);
    \item \emph{GPU SM Utilization} is the fraction of time, when 
    at least one warp was active on a multiprocessor, 
    averaged over all SMs;
    \item \emph{Bandwidth Utilization} is the measured network (PCIe / NVLink / RDMA) bandwidth.
\end{itemize}
We use NVIDIA's DCGM \cite{dcgm} to inspect the metrics of
device utilization on our testbeds.
\begin{figure*}[t]
\centering
\begin{minipage}[t]{0.325\linewidth}
\centering
\includegraphics[width=\linewidth]{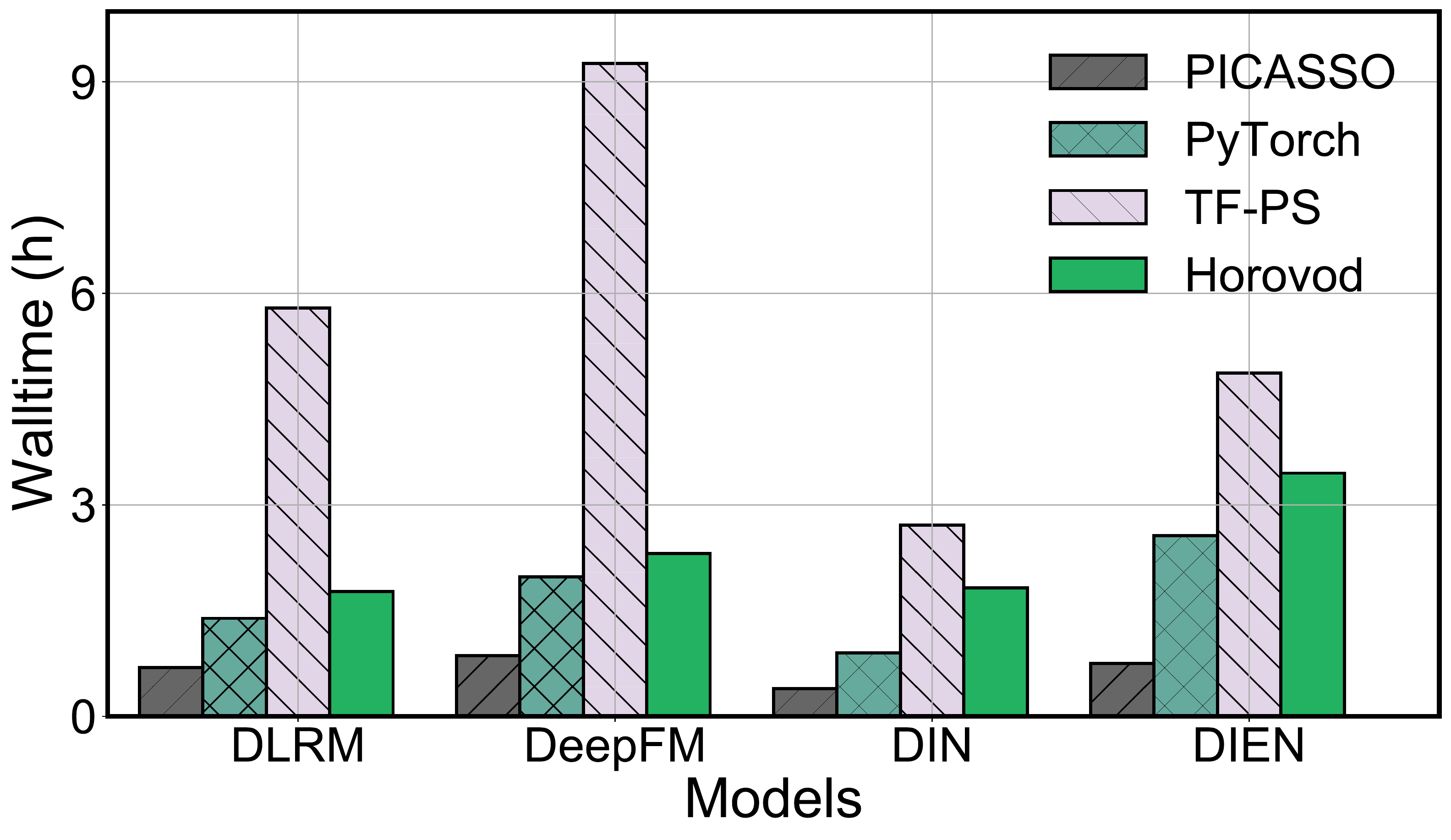}
\caption{Walltime in GPU core hours of training the four models completely by the compared training systems.}\label{fig:performance}
\end{minipage}
\hfill
\begin{minipage}[t]{0.325\linewidth}
\centering
\includegraphics[width=\linewidth]{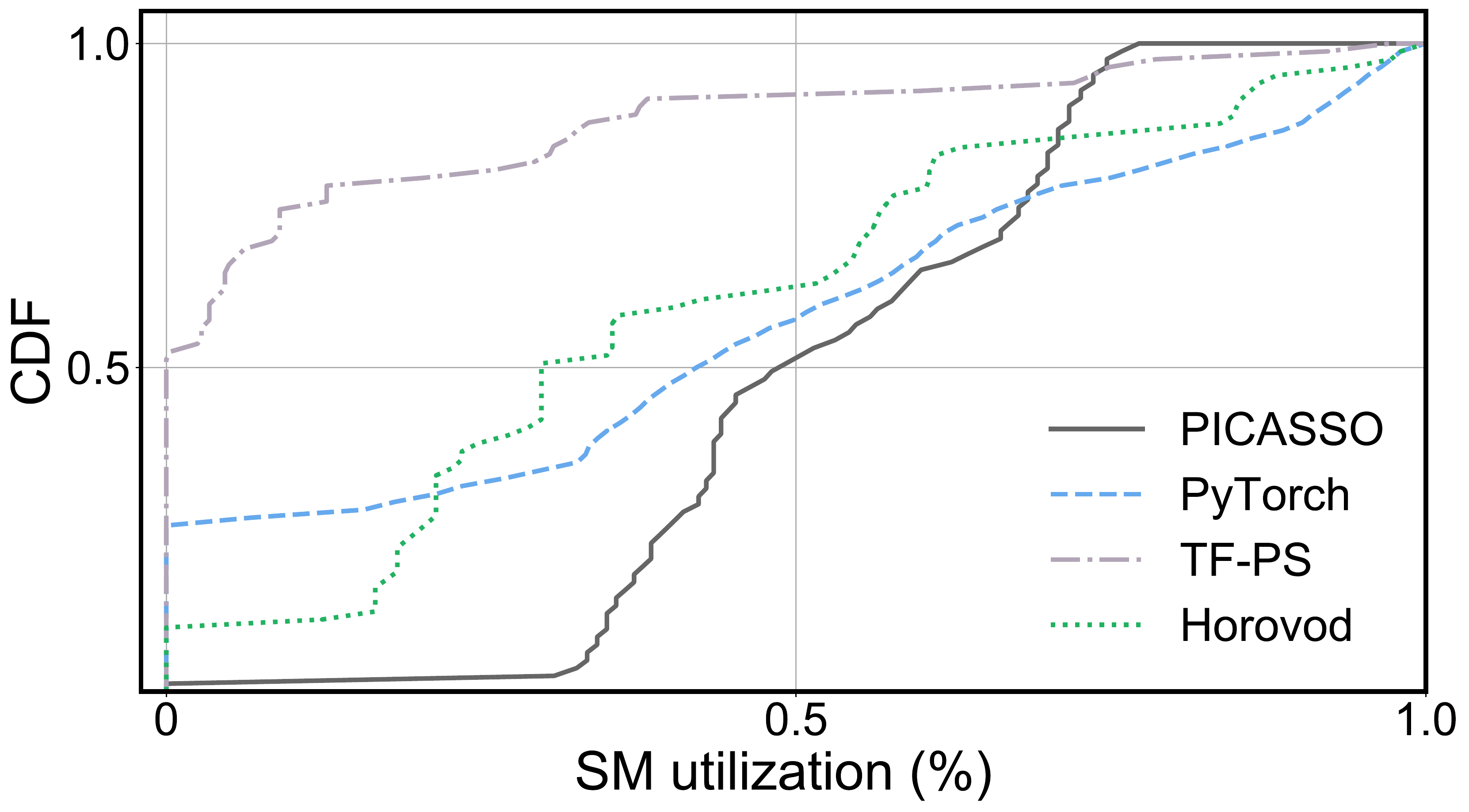}
\caption{CDF of SM utilization of training DLRM during the entire process by the four compared 
systems.}\label{fig:dlrm_sm}
\end{minipage}
\hfill
\begin{minipage}[t]{0.325\linewidth}
\centering
\includegraphics[width=\linewidth]{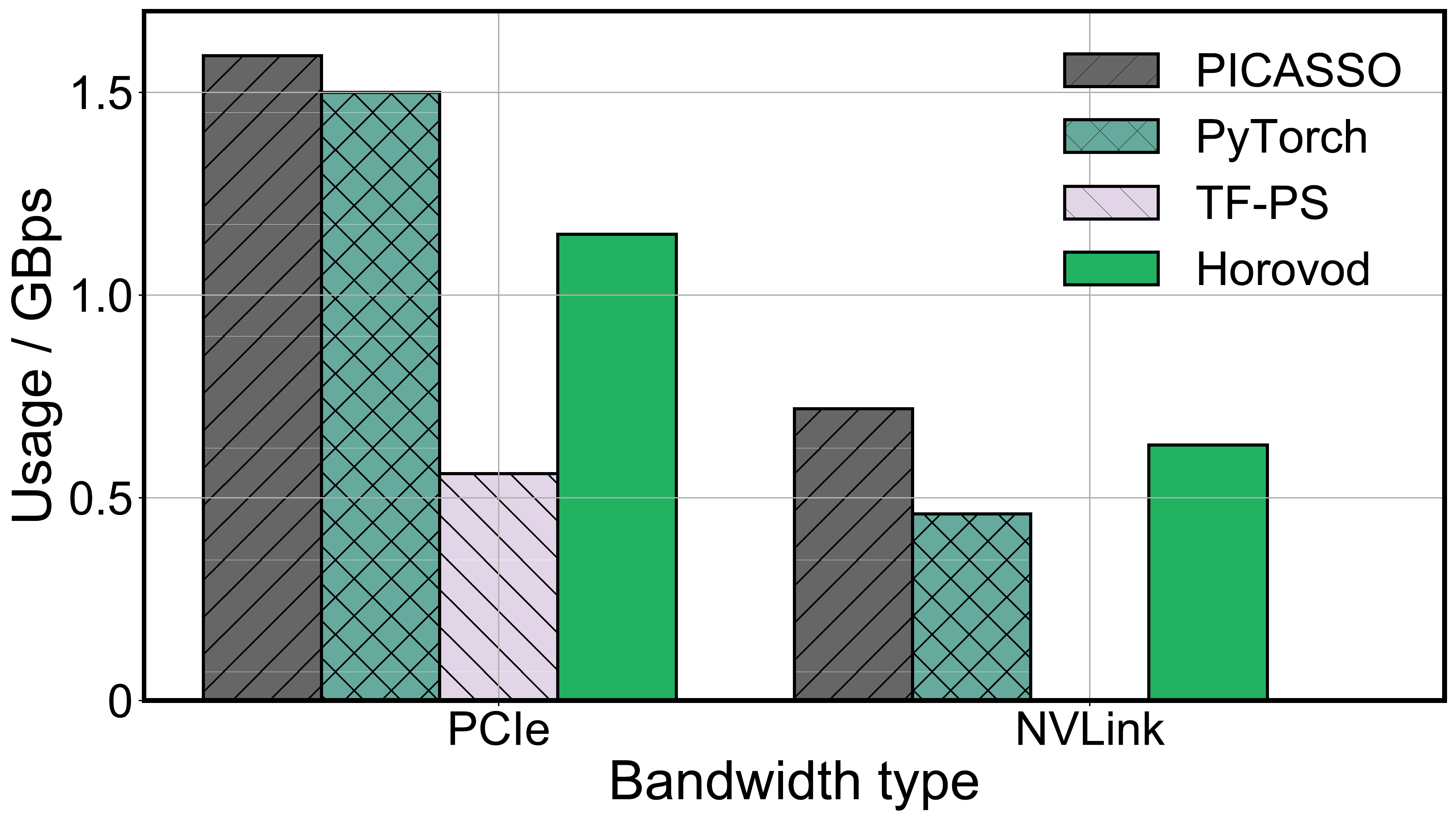}
\caption{PCIe and NVLink bandwidth consumption of training DLRM by the four compared
systems.}\label{fig:dlsd_pcie}
\end{minipage}
\end{figure*}

\subsection{Evaluation on Benchmarks (RQ1)}

We first examine the performance of \hbsys{} on one Gn6e cluster 
node over benchmarking tasks. We tune the batch sizes for each
framework's best throughput while maintaining the models' accuracy.
\begin{table}[tp]
    \centering
    \caption{AUC of trained models by four training systems}
    \label{tab:auc}
    \resizebox{\linewidth}{!}{
    \begin{tabular}{m{1cm}|m{1.6cm}m{1.6cm}m{1.6cm}m{1.6cm}}
    \toprule
     & \textbf{\hbsys{}} & \textbf{PyTorch} & \textbf{TF-PS} & \textbf{Horovod} \\
    \midrule
    DLRM & 0.8025 (42K) & 0.8025 (7K) & 0.8024 (6K) & 0.8025 (10K) \\
    DeepFM & 0.8007 (30K) & 0.8007 (7K) & 0.8007 (7K) & 0.8007 (8K) \\
    DIN & 0.6331 (32K) & 0.6329 (20K) & 0.6327 (16K) & 0.6329 (24K) \\
    DIEN & 0.6345 (32K) & 0.6344 (16K) & 0.6340 (12K) & 0.6343 (24K) \\
    \bottomrule
    \end{tabular}
    }
\end{table}

\noindent\textbf{Accuracy and Throughput}.
The AUCs, with corresponding batch sizes per GPU device, are listed in 
Tab.~\ref{tab:auc}. For DLRM and DeepFM, \hbsys{} achieves the same 
AUC with PyTorch and Horovod, which is better than the asynchronous
training by TF-PS. For DIN and DIEN, \hbsys{} even obtains a slightly 
improved accuracy than the others, which is instructive for industrial practice. \par

In terms of throughput, Fig.~\ref{fig:performance} 
records the training walltime of models by the four frameworks. 
TF-PS has the worst performance among the four
because of extensive data exchange and PCIe congestion between server and worker nodes. 
Horovod and PyTorch have much-improved performance over TF-PS due to the usage of collective 
communication primitives (i.e., Allreduce and AllToAll).
\hbsys{} presents the best performance, and the advantage is more 
remarkable on DIN and DIEN due to the relatively complicated workload
patterns (a hybrid of memory-, and computation-intensive layers;  and Alibaba dataset has a higher sparsity than Criteo).
The result shows that \hbsys{} impressively accelerates the training
by at least 1.9$\times$, and up to 10$\times$ compared 
to the baseline framework (TF-PS).

\noindent\textbf{Hardware Utilization}.
We then investigate the runtime utilization of the underlying hardware 
when training the DLRM model, and we plot the GPU SM utilization and
NVLink/PCIe bandwidth consumption in a 10-millisecond granularity in 
Fig.~\ref{fig:dlrm_sm} and Fig.~\ref{fig:dlsd_pcie}.
Although the other frameworks optimize some phases of training
(e.g., PyTorch and Horovod present intermittent 
high GPU SM utilization), they suffer from certain bottlenecks implied 
from the large CDF area of low GPU SM utilization.
In contrast,
\hbsys{} has barely any area of low GPU SM utilization, meaning that the 
bottlenecks on this testbed are effectively addressed by the software system optimization. 
In terms of bandwidth utilization, \hbsys{} is much better than the TF-PS baseline as leveraging the collective communication
primitives and the hardware coherency via NVLink. When compared to Horovod
and PyTorch, \hbsys{} still slightly improves bandwidth usage due
to pipelines in interleaving. 
The improvement of hardware utilization indicates that though the generic frameworks have pushed the computation efficiency to a peak level, \hbsys{} still succeeds to unleash the potential of the underlying hardware resources.

\begin{figure*}[htbp]
\centering
\begin{minipage}[t]{0.325\linewidth}
\centering
\includegraphics[width=0.99\linewidth]{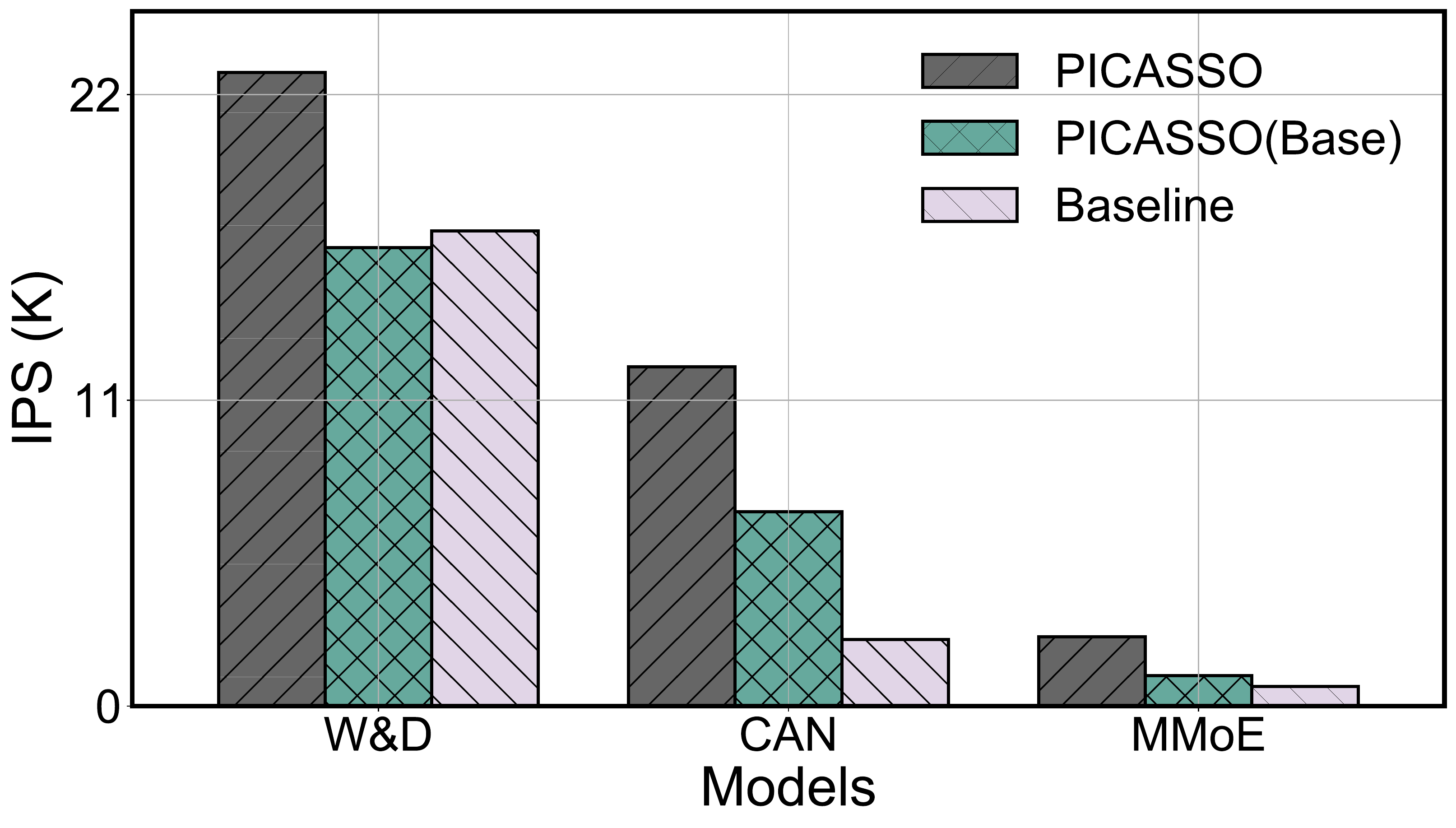}
\caption{Training performance of the three models by the three training systems in the in-house cluster.}\label{fig:acceleration}
\end{minipage}
\hfill
\begin{minipage}[t]{0.32\linewidth}
\centering
\includegraphics[width=\linewidth]{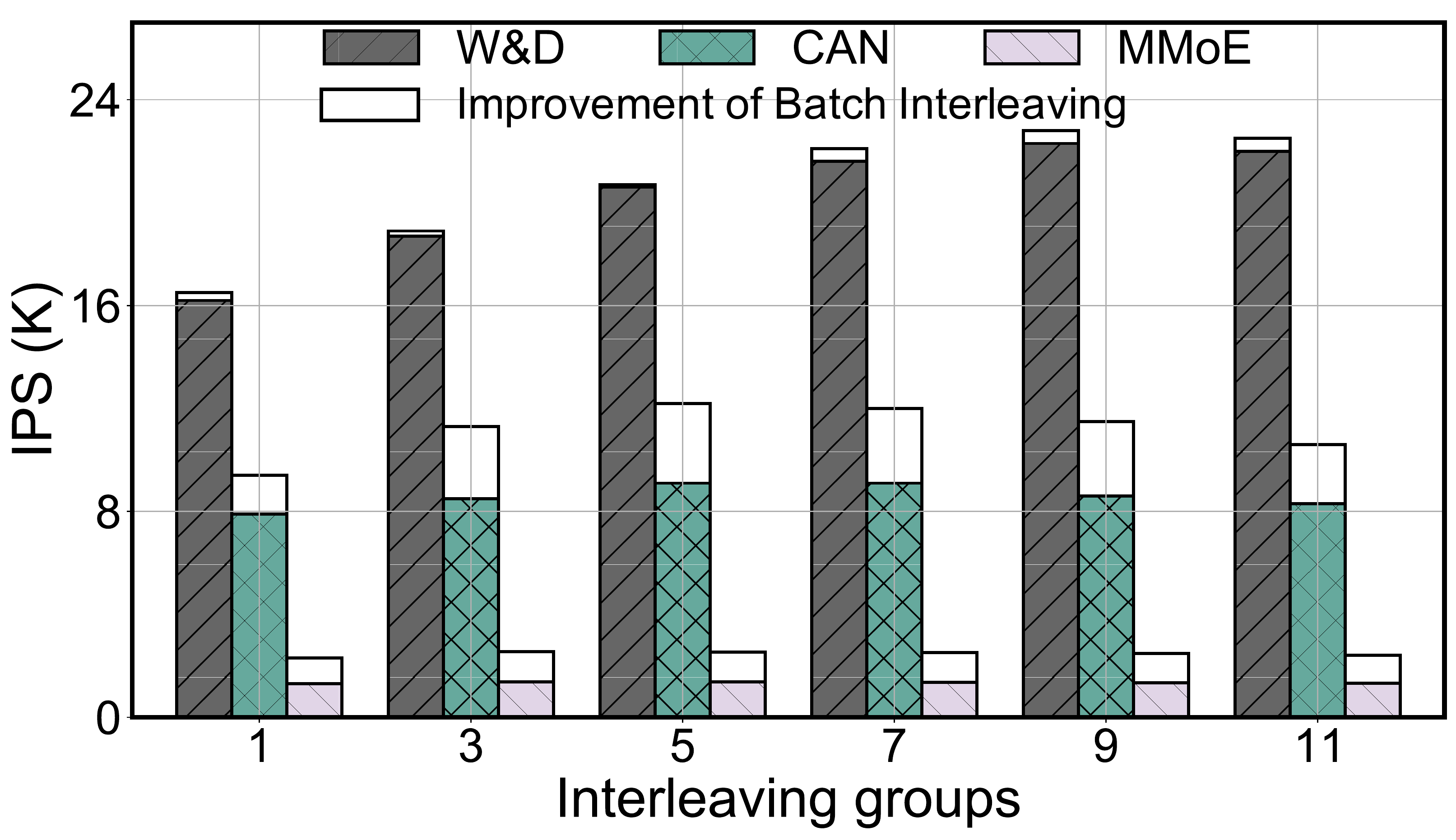}
\caption{Training performance by using 1 to 11 interleaving groups. The three models own 16, 19, 11 packed embedding respectively.}\label{fig:scale_up}
\end{minipage}
\hfill
\begin{minipage}[t]{0.32\linewidth}
\centering
\includegraphics[width=\linewidth]{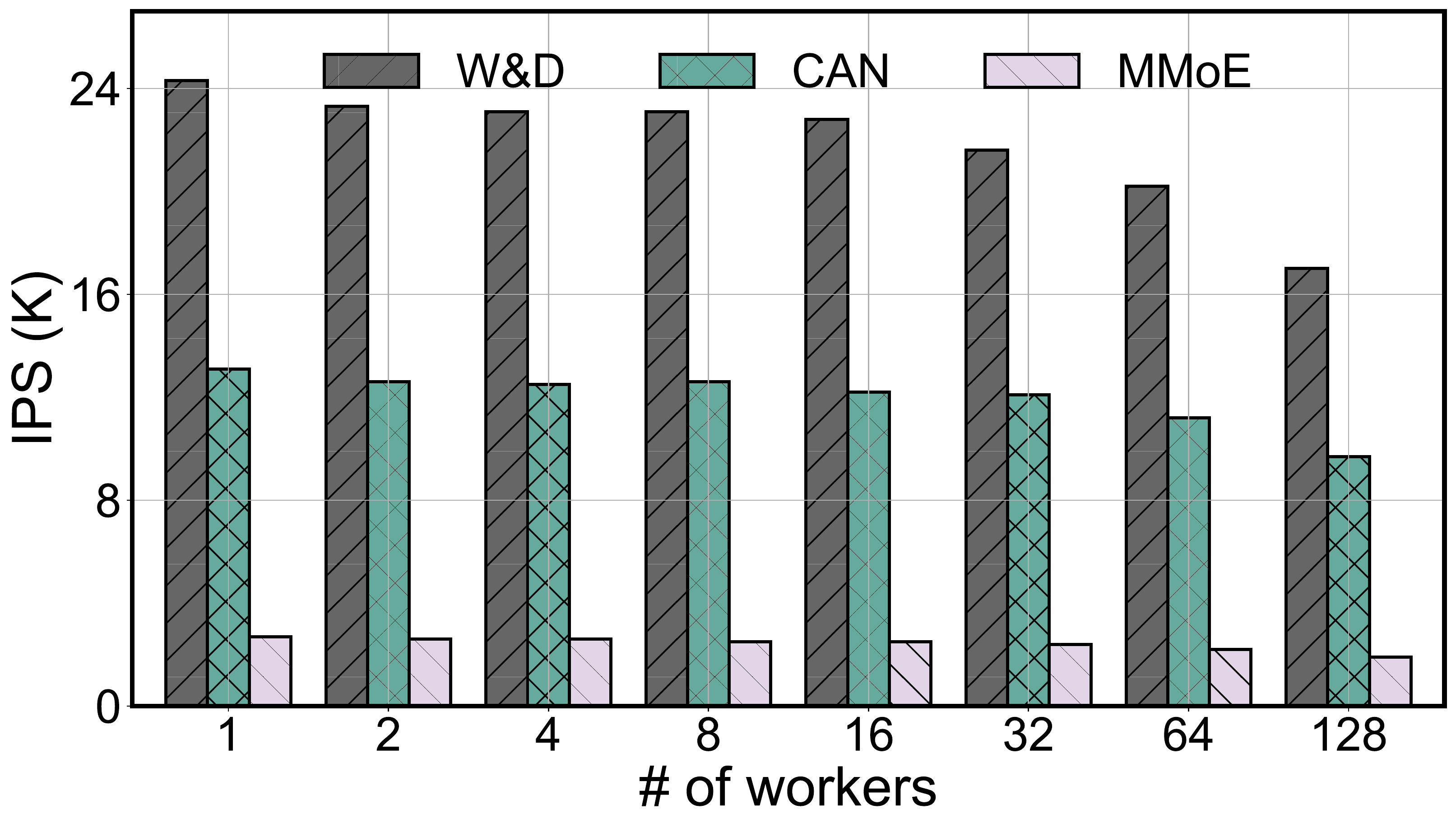}
\caption{Training performance of scaling-out of \hbsys{} ranging from 1 to 128 workers in the in-house cluster.}
\label{fig:scale_out}
\end{minipage}
\end{figure*}

\subsection{Evaluation of System Design (RQ2)}

\begin{table}[t]
    \centering
    \caption{Ablation study on \hbsys{}}
    \label{tab:ablation_study}
    \resizebox{\linewidth}{!}{
    \begin{tabular}{m{0.7cm}m{2cm}|m{0.8cm}m{1.1cm}m{1.1cm}m{1.2cm}}
    \toprule
     & & \textbf{IPS} & \textbf{PCIe (GBps)} & \textbf{Comm. (Gbps)} & \textbf{SM util. (\%)} \\
    \midrule
    \multirow{4}{0.7cm}{W\&D} & \hbsys{} & 22,825 & 1.57 & 2.48 & 32 \\
    & \, w/o Packing & 17,827 & 1.54 & 1.84 & 23 \\
    & \, w/o Interleaving & 16,218 & 1.49 & 1.69 & 21 \\
    & \, w/o Caching & 19,264 & 1.51 & 2.07 & 25 \\
    \midrule
    \multirow{4}{0.7cm}{CAN} & \hbsys{} & 12,218 & 2.59 & 8.50 & 62 \\
    & \, w/o Packing & 8,769 & 2.55 & 6.66 & 45 \\
    & \, w/o Interleaving & 7,957 & 2.02 & 6.94 & 43 \\
    & \, w/o Caching & 10,829 & 2.60 & 7.41 & 51 \\
    \midrule
    \multirow{4}{0.7cm}{MMoE} & \hbsys{} & 2,546 & 2.31 & 6.61 & 98 \\
    & \, w/o Packing & 2,270 & 2.27 & 6.10 & 96 \\
    & \, w/o Interleaving & 1,319 & 1.87 & 3.80 & 64 \\
    & \, w/o Caching & 2,401 & 2.28 & 6.44 & 98 \\
    \bottomrule
    \end{tabular}
    }
\end{table}

Industrial \hsdl{} workloads are usually much more complicated than the benchmarking workloads, regarding both model architecture and data distribution.
We investigate the effectiveness of \hbsys{} in industry services by W\&D, CAN, and MMoE models over
industrial datasets.
Meanwhile, we measure the fine-grained contribution of packing, interleaving and caching
by training throughput.
The evaluation is conducted over 16 nodes in the
EFLOPS cluster if not specified explicitly.
We use the commonly-used asynchronous PS strategy of Tensorflow 
at Alibaba as the baseline.
We also implement \hbsys{} without
software system optimization, denoted as ``\hbsys{}(Base)'', which can be seen as a pure hybrid-parallel training strategy framework.
Fig.~\ref{fig:acceleration} depicts the IPS, where we
observe a 4$\times$ acceleration on CAN and MMoE.
We then dive into the software system optimizations via an ablation study.

\noindent\textbf{Ablation Study}.
We remove the software system optimization from \hbsys{},
in turn, to verify its effect on \hsdl{} tasks and 
collect metrics in Tab.~\ref{tab:ablation_study}.

By using the packing approach, the fragmentary operations on 
feature embedding are packed together, leading
to an improvement on IPS by 30\%
and the correspondingly increased hardware usage of PCIe, network and GPU SMs.
The interleaving approach utilizes pipelines 
to hide memory access and network latency by computation-intensive operations. 
Obviously, MMoE owns the most
complicated computation workload among the three models and thus benefits most from this optimization.
The interleaving approach significantly raises the performance of MMoE by 93\%. This is consistent with
the analysis that the two models suffer heavy computation load, and \hbsys{} helps diffuse the pulse-like GPU usage throughout the entire training process.
W\&D, not having sufficient computation, would benefit from the data-interleaving to mitigate the congestion on PCIe and network.
Caching relies on leveraging the distribution 
of input data. Thus, we run 100 steps as a warm-up to collect the
statistics and then set the Hot-storage size to 1GB on GPU memory to maintain the unique IDs' hit ratio above 20\% within each batch.
HybridHash improves the performance by up to 13\%,
which attributes to a balanced utilization of PCIe and GPUs.


\begin{table}[t]
    \centering
    \caption{Number of operations in computational graph}
    \label{tab:nodes}
    \begin{tabular}{c|cccc}
    \toprule
    \multirow{2}{*}{\textbf{Model}} & \multicolumn{2}{c}{\textbf{\# of operations}} &  \multicolumn{2}{c}{\textbf{\# of packed embedding}} \\
     & Baseline & \hbsys{} & Baseline & \hbsys{} \\
    \midrule
    W\&D & 100,039 & 14,882 (14.9\%)  & 204 & 16 (7.8\%) \\
    CAN & 381,364 & 67,985 (17.8\%) & 364 & 19 (5.2\%) \\
    MMoE & 300,524 & 75,217 (25.0\%) & 94 & 11 (11.7\%) \\
    \bottomrule
    \end{tabular}
\end{table}

\noindent\textbf{Effectiveness of Packing}.
\hsdl{} models tend to own fragmentary operations for the multi-field embedding and feature interactions. 
We dump the computational graph of the three models through Baseline and \hbsys{}, and the number of operations and packed embedding are shown in Tab.~\ref{tab:nodes}.
The statistic implies that \hbsys{} dramatically reduces the fragmentary operations, even if the interleaving optimization supplements a certain amount of operations to pipeline the executions.

\noindent\textbf{Interleaving Groups}.
The number of interleaving groups affects the efficiency of the embedding layer.
Fig.~\ref{fig:scale_up} shows the throughput by varying the number of K-interleaving groups.
Obviously, the communication-intensive workloads, i.e., W\&D and CAN, can benefit from the increased interleaving as the combination of packed embedding uniformizes the usage of each hardware resource.
We also see that the batch interleaving contributes differently to the models.
The result reflects that utilizing more micro-batches would greatly improve the performance of the computation-intensive workload, i.e., CAN and MMoE, by meeting the saturation of GPU.
It reveals that the interleaving strategies are effective for WDL workloads when there is sufficient input data and underutilized hardware.

\begin{table}[t]
    \centering
    \caption{Hit ratio and IPS by varying the size of Hot-storage}
    \label{tab:cache}
    \resizebox{\linewidth}{!}{
    \begin{tabular}{m{1.3cm}|m{1.0cm}m{0.8cm}m{1.0cm}m{0.8cm}m{1.0cm}m{0.8cm}}
    \toprule
    \multirow{2}{*}{\textbf{Hot-Storage}} & \multicolumn{2}{c}{\textbf{W\&D}} & \multicolumn{2}{c}{\textbf{CAN}} & \multicolumn{2}{c}{\textbf{MMoE}} \\
     & Hit ratio & IPS  & Hit ratio & IPS  & Hit ratio & IPS \\
    \midrule
    256MB & 9\% & -11\% & 20\% &	-19\% & 9\% &	-3\% \\
    512MB & 18\% &	-5\% & 28\% &	-10\% & 16\% &	-1\% \\
    1GB & 24\% &	+0\% & 37\% &	+0\% & 21\%	& +0\%  \\
    2GB & 28\% &	+1\% & 44\% &	+5\% & 24\% &	+0\% \\
    4GB & 31\% &	-3\% & 45\% &	+2\% & 27\% &	-2\% \\
    \bottomrule
    \end{tabular}
    }
\end{table}
\noindent\textbf{Size of Hot-storage}.
In the industrial \hsdl{} workload, it is impossible to foresee the size of the embedding tables since the model should constantly deal with the newly-emerged categorical feature IDs.
We set the size of Hot-storage to 1GB in the previous evaluations to ensure 20\% hit ratio.
Tab.~\ref{tab:cache} depicts the cache hit ratio and improvement of IPS by varying the size of Hot-storage.
Larger cache size carries more embedding, yet we find an apparent marginal effect of the hit ratio when the cache size reaches above 2GB.
Although the large cache hits more ID queries, the occupation of GPU memory forces the training to compromise the batch size, leading to a slight reduction in the overall throughput.
Hence, it is no need to pursue high cache hit ratio by setting an excessively large cache size in \hsdl{} workload.

\noindent\textbf{Scaling Out}.
We scale out the training clusters from one \hbworker{} to 128 \hbworker{}s and 
illustrate the performance by IPS in Fig.~\ref{fig:scale_out}.
The correlation between IPS and the number of \hbworker{}s shows that 
\hbsys{} achieves near-linear scalability on CAN and MMoE while
attaining a sublinear throughput on W\&D.
This result implies that \hbsys{} can amortize the additional communication overhead
from the increasing number of \hbworker{}s and handle large-scale \hsdl{} training.

\subsection{Applicability Experiment (RQ3)}~\label{sec:rq3}

\begin{table}[h]
    \centering
    \caption{Training throughput of 12 AUC prediction models by \hbsys{} and in-house XDL.}
    \label{tab:performance}
    \resizebox{\linewidth}{!}{
    \begin{tabular}{m{1.8cm}|m{2.8cm}m{2.4cm}m{1.9cm}}
    \toprule
     \textbf{Model} & \textbf{Batch size} & \textbf{GPU SM utilization} & \textbf{IPS} \\
    \midrule
	LR~\cite{gai2017learning} & 20K $\xrightarrow{}$ 36K (20K x 2) & 9 $\xrightarrow{}$ 22 (+144\%) & 12.0K $\xrightarrow{}$ 25.9K (+115\%) \\
	W\&D~\cite{cheng2016wide} & 19K $\xrightarrow{}$ 36K (18K x 2) & 21 $\xrightarrow{}$ 35 (+67\%) & 14.7K $\xrightarrow{}$ 22.2K (+50\%) \\
	TwoTowerDNN\cite{fan2019mobius} & 12K $\xrightarrow{}$ 36K (12K x 3) & 35 $\xrightarrow{}$ 97 (+177\%) & 4.7K $\xrightarrow{}$ 12.1K (+160\%) \\
	DLRM~\cite{mudigere2021high} & 10K $\xrightarrow{}$ 30K (10K x 3) & 38 $\xrightarrow{}$ 98 (+158\%) & 3.8K $\xrightarrow{}$ 10.4K (+171\%) \\
	DCN~\cite{wang2017deep} & 14K $\xrightarrow{}$ 36K (12K x 3) & 56 $\xrightarrow{}$ 92 (+64\%) & 9.0K $\xrightarrow{}$ 13.7K (+52\%) \\
	xDeepFM~\cite{lian2018xdeepfm} & 6K $\xrightarrow{}$ 20K (5K x 4) & 45 $\xrightarrow{}$ 98 (+117\%) & 3.1K $\xrightarrow{}$ 5.9K (+89\%) \\
	ATBRG~\cite{feng2020atbrg} & 3K $\xrightarrow{}$ 6K (3K x 2) & 13 $\xrightarrow{}$ 26 (+100\%) & 0.8K $\xrightarrow{}$ 1.4K (+82\%) \\
	DIN~\cite{zhou2018deep} & 15K $\xrightarrow{}$ 45K (15K x 3) & 34 $\xrightarrow{}$ 80 (+135\%) & 7.5K $\xrightarrow{}$ 16.0K (+113\%) \\
	DIEN~\cite{zhou2019deep} & 15K $\xrightarrow{}$ 45K (15K x 3) & 29 $\xrightarrow{}$ 75 (+159\%) & 7.3K $\xrightarrow{}$ 15.6K (+115\%) \\
	DSIN~\cite{feng2019deep} & 9K $\xrightarrow{}$ 27K (9K x 3) & 40 $\xrightarrow{}$ 93 (+133\%) & 4.7K $\xrightarrow{}$ 9.8K (+111\%) \\
	CAN~\cite{zhou2020can} & 12K $\xrightarrow{}$ 48K (12K x 4) & 17 $\xrightarrow{}$ 75 (+341\%) & 3.9K $\xrightarrow{}$ 12.1K (+210\%) \\
	STAR~\cite{sheng2021one} & 2K $\xrightarrow{}$ 8K (2K x 4) & 32 $\xrightarrow{}$ 98 (+206\%) & 0.6K $\xrightarrow{}$ 2.0K (+215\%) \\
    \bottomrule
    \end{tabular}
    }
\end{table}

\noindent\textbf{Varying Feature Interactions}. 
We further investigate the performance of \hbsys{} on more industrial-scale \hsdl{} models with various types of feature interaction modules.
We select 12 AUC prediction models, and tune the hyperparameters to ensure the convergence of the models.
The models are slightly modified to cope with the Product-2 dataset.
To show the adaptability of \hbsys{}, we take the in-house optimized XDL as the baseline to train those models in synchronous PS training mode.
The performance of the 12 models is listed in Table~\ref{tab:performance}.
Obviously, the proposed \hbsys{} significantly improves the GPU utilization compared to the in-house optimized XDL~\cite{jiang2019xdl} framework.
It indicates that with the software system optimization, \hbsys{} is able to be aware of the hardware bound, and unleash the hardware potential for various \hsdl{} model architectures.

\begin{table}[t]
    \centering
    \caption{Performance of CAN by varying number of feature fields on synthetic dataset}
    \label{tab:massive_field}
    \resizebox{\linewidth}{!}{
    \begin{tabular}{m{1.6cm}|m{0.7cm}m{0.7cm}m{0.7cm}m{0.7cm}m{0.84cm}m{0.7cm}m{0.84cm}m{0.84cm}}
    \toprule
    Feature field & 1 & 2 & 3 & 4 & 5 & 6 & 7 & 8 \\
    \midrule
    \hbsys{} & 12.20 &	6.14 &	4.13 &	3.13 &	2.50 &	2.09 &	1.82 &	1.61 \\
    AP & 12.20 &	6.10 &	4.07 &	3.05 &	2.44 &	2.03 &	1.74 &	1.53 \\
    Increment & 0.0\% &	+0.6\% & +1.7\% &	+2.5\% &	+2.6\% &	+2.7\% &	+4.3\% &	+5.3\% \\
    XDL & 2.40 &	1.18 &	0.75 &	0.56 &	0.42 &	0.36 &	0.31 &	0.25 \\
    AP & 2.40 &	1.20 &	0.80 &	0.60 &	0.48 &	0.40 & 	0.34 &	0.30 \\
    Increment & 0.0\% &	-1.5\% &	-6.8\% &	-6.1\% &	-13.1\% &	-9.6\% &	-10.3\% &	-15.3\% \\
    \bottomrule
    \end{tabular}
    }
\end{table}

\noindent\textbf{Increasing Feature Fields}.
We present the performance of PICASSO by varying the number of feature fields. 
As currently we do not operate business that ingests thousands of feature fields, we make a synthetic dataset by duplicating the feature fields of Product-2. 
Therefore, the number of feature fields becomes an integral multiple of 364 in the data source. 
Correspondingly, we duplicate the feature interaction layers to process the synthetic dataset. 
The outputs of the feature interaction layers are concatenated together to feed to a mutual MLP. 
Tab.~\ref{tab:massive_field} depicts the IPS by PICASSO and the in-house optimized XDL, as well as the theoretical IPS by arithmetic progression (AP). 
Though the requirements of the underlying hardware resources increase by the number of feature fields, PICASSO performs slightly better than AP due to the packing of the fragmentary operations. 
In contrast, the PS baseline suffers from the massive operations by the large number of feature fields and the constituent feature interactions, thereby presenting significant IPS drop compared to AP.

\noindent\textbf{Discussion}.
Evaluations show that \hbsys{} successfully unleashes the potential of hardware resources, where \hbsys{} uniformizes the hardware usage for high overall hardware utilization (RQ2) and provides diversified optimization for \hsdl{} models with different attributes (RQ3).
Employing super-large batch size in training requires certain auxiliary approaches (e.g., global batch normalization \cite{chen2020simple}, Lamb optimizer \cite{you2019large}), which can be applied in \hbsys{} through implementation in Tensorflow.
Obviously, should we deploy customized hardware to enhance specific hardware resources, or if particular \hsdl{} tasks presented high tolerance for accuracy loss from the precision-lossy mechanisms, the performance of \hbsys{} will surely be further improved.

\section{Deployment in Production}
\begin{table}[t]
    \centering
    \caption{Performance of \hbsys{} at Alibaba Cloud}
    \label{tab:online_performance}
    \resizebox{\linewidth}{!}{
    \begin{tabular}{m{1.6cm}|m{1.6cm}m{1.2cm}m{2.5cm}}
    \toprule
     & \textbf{Average task walltime (h)} & \textbf{GPU SM util. (\%)} & \textbf{Bandwidth (Gbps)} \\
    \midrule
    XDL & 8.6 & 15 & 1.412 (TCP) \\
    \hbsys & 1.4 & 75 & 6.851 (TCP+RDMA) \\
    \bottomrule
    \end{tabular}
    }
\end{table}
We implement \hbsys{} on top of Tensorflow.
It has been deployed in our on-premises clusters since October 2020 to 
serve business of online and offline \hsdl{} tasks, including information retrieval, advertisement bidding, recommendation, and search ranking.
Since we deployed \hbsys{}, training throughput has been greatly improved, and impressive performance has been achieved in a number of highly-promoted sales.
In-house sophisticated scheduling and 
failover-recovery strategies are employed by \hbsys{} for robust
training \cite{huang2019yugong,feng2021scaling}, which are beyond this paper's scope. 
\par


The arithmetic computation is still very heavy in training state-of-the-art \dfi{} designs. 
Benefiting from popular CV and NLP training acceleration approaches, we have applied the latest solutions, such as GPU acceleration libraries (e.g., CUTLASS \cite{cutlass} and CuDNN \cite{cudnn}), operator-level graph replacement \cite{chen2018tvm}, compiler optimization \cite{dean2017machine}, and quantitative communication\cite{jiang2018linear} to \hbsys{}.
We also implement topology-aware communication\cite{dong2021accl} to avoid IO tasks on GPU devices from the same node competing for limited NIC resources.
These accelerations are orthogonal to the optimization of \hbsys{}. 
We provide users with a flexible interface to invoke these methods when tuning their design. 
Other emerging technologies can be integrated into \hbsys{} via the Tensorflow ecosystem.

We instrument a production training cluster of one Tesla-V100 per worker, 
which runs hundreds of daily \hsdl{} workloads. These workloads present remarkably 
different training intensities by various input features, embedding dimensions,
feature interaction modules, and shapes of MLP.
We record job statistics of the succeeded training tasks from
Jun. 1st, 2021, to Nov. 15th, 2021.
To make a comparison, we prepare previously deployed XDL \cite{jiang2019xdl} in
another production cluster with comparable types of workloads.
The results in Tab.~\ref{tab:online_performance} show that \hbsys{} brings around 6$\times$ performance acceleration on average and contributes to improving
the utilization of underlying hardware.
The throughput acceleration reduces the delay of daily continuous delivery by 7 hours on average.
We further probe into several representative models (with entirely different model 
architecture and data distribution) from the monitored cluster and present the 
required walltime over 128 Tesla-V100s to train petabyte-scale data accumulated by one year as shown in
Tab.~\ref{tab:year}.
The statistics reveal that \hbsys{} reduces model training of 100-billion-scale parameters from one month to 2 days.
Moreover, regarding a \hsdl{} model with 1-trillion-scale parameters (one of the currently largest models satisfying
real-time inference throughput demand in our business),
the training completes within nine days, while the baseline framework is estimated to occupy the resources for more than three months.
This training acceleration is 
critical in the latest ML/DL trends of providing high \hsdl{} business value.

\section{Related Work}
\label{sec:related_work}

We summarize the trending research approaches with respect to training \hsdl{} jobs into three categories as follows:

\noindent \textbf{Hardware Customization}. For a specific \hsdl{} workload with a high business value, it is profitable to customize the hardware itself to achieve a
cutting-edge performance and throughput. AIBox/PaddleBox \cite{zhao2019aibox,zhao2020distributed} leverages non-volatile memory to drastically reduce the training scale from an MPI cluster
with hundreds of CPUs to a single machine with 8 GPUs. HugeCTR/Merlin is a customized framework running on NVIDIA's DGX-1/DGX-2 supernodes equipped with high-end 
interconnects named NV-Switch. Zion \cite{smelyanskiy2019zion} and RecSpeed \cite{krishna2020accelerating} customize their node specification for DLRM~\cite{mudigere2021high} 
and its variants by adding more NICs and RoCEs to alleviate the I/O bottleneck. Nevertheless, hardware customization is still expensive and a waste of resources when facing rapid shifts in
\dfi{} designs. Further, training systems upon customized hardware are difficult to scale out on cloud elastically.   

\begin{table}[t]
    \centering
    \caption{Walltime (GPU-core hour) to train data accumulated by one year (``P'' indicates the predicted walltime)}
    \label{tab:year}
    \resizebox{\linewidth}{!}{
    \begin{tabular}{m{1.8cm}|m{1cm}m{1cm}m{1.6cm}m{1.6cm}}
    \toprule
    Model scale & $\sim$ 1B & $\sim$ 10B & $\sim$ 100B & $\sim$ 1T \\
    \midrule
    XDL & 2,072 & 11,013 & 88,129 (P) & 323,480 (P)\\
    \hbsys{} & 747 & 2,285 & 6,091 & 27,256 \\
    \bottomrule
    \end{tabular}
    }
\end{table}

\noindent \textbf{Subsystem Optimization}. Subsystem optimization diagnoses specific bottlenecks and improves the performance of certain workloads. 
For example, the communication protocol in BytePS \cite{peng2019generic} accelerates the 
data exchange in PS strategy. Kraken \cite{xie2020kraken} develops memory-efficient table structure to hold parameters of embedding layers. ScaleFreeCTR~\cite{guo2021scalefreectr}
utilizes GPU to accelerate the embedding lookup of parameters stored in DRAM. 
Het~\cite{miao2021het} introduces staleness to embedding update which is suitable for the \dfi{} designs with small-size local embedding tables.
These optimizations are likely to miss opportunities of improving overall performance systematically owing to the unawareness of either sparse manipulation or intensive computation of \dfi{}s, while precision-lossy operations like staleness would do harm to the E-commerce \hsdl{} models.

\noindent \textbf{Generic DNN Training Optimization}. There are already a variety of proposed training frameworks targeting dense models
from domains such as CV and NLP. These frameworks provide meticulous strategies for splitting and pipelining
workloads during training. Megatron \cite{shoeybi2019megatron} speeds up the transformer module in NLP workloads.  
GPipe \cite{huang2018gpipe} implements pipeline over mini batches, and Pipedream \cite{narayanan2019pipedream} further fills the bubble between forward and backward pass by weight stashing.
GShard~\cite{lepikhin2020gshard} relies on a compilation approach to shard parameters and activations.
Unfortunately, \hsdl{} models are usually sensitive to numeric precision and gradient staleness \cite{xu2021step,zhuang2021accumulated,cai2020rethinking}, and \hsdl{} workload has 
much more operators on dynamic shape of data than CV and NLP models. Hence, these generic DNN training optimizations may not apply to \hsdl{} workload at an industrial scale. 


\section{Conclusion}
\label{sec:conclusion}

In this paper, we introduce \hbsys{}, a deep learning training system upon Tensorflow, to accelerate the training of \hsdl{} models on commodity hardware with the awareness of model architecture and data distribution. 
With the investigation of representative workloads at Alibaba Cloud, we
design an advantageous training framework and provide 
workload-aware software optimization: 1) packing the embedding tables and the subsequent operations to reduce the fragmentary operations that are not friendly to GPU-centric training; 
2) interleaving the embedding layer and the feature interaction to diffuse the pulse-like usage throughout the entire training process;
3) caching the frequently-visited categorical IDs to expedite the repeated embedding queries.
Product deployment at Alibaba Cloud demonstrates that a 1-trillion
parameter \hsdl{} model through one-year petabyte-scale data can be efficiently trained
in 27,256 GPU core hours, significantly reducing the 
training cost by a factor of 12.
\hbsys{} helps decrease the delay of daily continuous delivery by 7 hours, which is crucially important for state-of-the-art recommender systems.

\section*{Acknowledgement}
This work is supported by Alibaba Group. 
We thank Lixue Xia, Wencong Xiao, Shiru Ren, Zhen Zheng, and Zheng Cao for useful pointers regarding the writing of this paper.
We appreciate Hua Zong, Kaixv Ren, Wenchao Wang, Xiaoli Liu, Yunxin Zhou, Hao Li, Guowang Zhang, Sui Huang and Lingling Jin for implementing the key components of the training and serving infrastructure.

\bibliographystyle{IEEEtran}
\bibliography{santa}

\end{document}